Leibniz University Hannover, Germany
L3S Research Center, Hannover

# Capturing, Documenting and Visualizing Search Contexts for building Multimedia Corpora

A PROJECT REPORT

Submitted by

**Zeon Trevor Fernando (10BCE1113)**

*In partial fulfillment for the award*

of

## Bachelor of Technology

Degree in

## Computer Science and Engineering

## School of Computing Science and Engineering

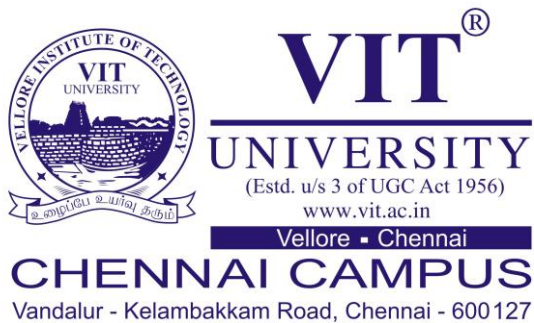
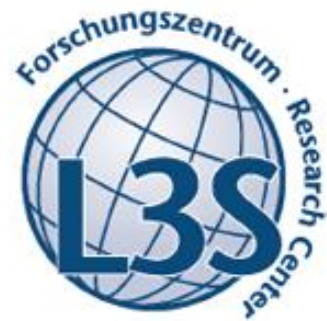

**July - 2014**

# CONTENTS

# LIST OF TABLES

# LIST OF FIGURES

# LIST OF ABBREVIATIONS

| Abbreviation | Expansion |
|---|---|
| JSF | Java Server Faces |
| MVC | Model View Controller |
| REST | Representational State Transfer |
| API | Application Programming Interface |
| XML | Extensible Markup Language |
| URL | Uniform Resource Locator |
| IDE | Integrated Development Environment |
| GUI | Graphical User Interface |
| HTTP | Hypertext Transfer Protocol |
| JSON | JavaScript Object Notation |
| JAXB | Java Architecture for XML Binding |
| URI | Uniform Resource Identifier |
| RIA | Rich Internet Application |
| UI | User Interface |
| CSS | Cascading Style Sheets |
| SDK | Software Development Kit |
| POJO | Plain Old Java Objects |
| MIME | Multipurpose Internet Mail Extensions |

# ABSTRACT

In Social Science research, multimedia documents are often collected to answer particular research questions like: "Which of the aesthetic properties of a photo are considered important on the web" or "How has Street Art developed over the past 50 years". Therefore, a researcher generally issues multiple queries to a number of search engines. This activity may span over long time intervals and results in a collection which can be further analyzed. Documenting the collection building process which includes the context of the carried out searches is imperative for social scientists to reproduce their research. Such context documentation consists of several user actions and search attributes like: the issued queries; the results clicked and saved; duration a particular result was viewed for; the set of results that was displayed but neither clicked, nor saved; as well as user annotations like comments or tags.

In this thesis I will describe a search process tracking module and a search history visualization module. These modules can be integrated into keyword based search systems through a REST API I developed to help capture, document and revisit past search contexts while building a web corpora. Finally, I detail the implementation of how my work was integrated into the LearnWeb2.0 platform - a multimedia web2.0 search and sharing application which can obtain resources from various web2.0 tools such as Youtube, Bing, Flickr, etc using keyword search.

# CHAPTER 1

# INTRODUCTION

## 1.1 General

According to the classical model of information retrieval a user uses a search engine for the purpose of satisfying an information need which is associated with some task. But it is shown that a web search is only informational less than 50% of the time and the rest could be navigational (to find a site you want to visit) or transactional (redirect to a site for transactions like shopping, downloading, etc) [1]. The focus of my thesis is implementing a system to support information seeking tasks and services to capture the context of these tasks which could help build multimedia corpora. Users express their needs as a query and scan through the results returned, in some cases their needs would be satisfied when the answer is obtained in the first page of results such services are provided by Google's oneboxes [2] and Yahoo's shortcut [3]. In other cases their needs are partially satisfied this happens when the information need of the user is complex, that is, a single web page cannot provide all the information. So in such cases user clicks on a few results returned and explores further, by refining their queries based on the information gathered to complete the entire task. These sessions generally span a few minutes to several days, examples include planning for a holiday trip (searching for accommodation, travel, places of attraction), education purposes (completing an assignment or researching on admissions) or building multimedia corpora for social science research. Users generally interleave multiple tasks during a search session or there could be task switching due to external interruptions, thus there needs to be modules which will assist in resuming these tasks at a later time. This highlights the need for modules that could support the second case, which keeps track of the search context that is the queries issued and the useful results found for a particular search and ways to re-access this information in order to reconstruct the context of that previous task.

Web services are a collection of open protocols and standards used for exchange of data between applications or systems. Web based applications implemented in various programming languages or frameworks and deployed on different platforms can use web services to exchange data over a network. One major class of web services is *REST*-compliant web services; the primary purpose of these services is to manipulate XML representations of web resources using a uniform set of stateless operations [4]. Representation state transfer (REST) is an architectural style for developing web services and applications in which agents provide uniform interface functionality such as create, retrieve, update and delete rather than application specific interfaces. Resources are manipulated only by the exchange of representations and the interactions between components are stateless, that is, the meaning of a message does not depend on the state of the conversation. Due to its uniform interface constraint, it helps decouple the client from the server thus enabling each part to evolve independently. Web applications have moved away from Simple Object Based Protocol (SOAP) based web services towards RESTful web services due to the ease of use and non-requirement of XML based web service protocols to support their interfaces. Thus in order to capture the search context and to re-access this search context information I built a RESTful web service so that it can be easily integrated into keyword search applications with only the need to develop a corresponding client for that application in order to send requests to the service.

The keyword search application we considered for our scenario is LearnWeb2.0, which is a system that integrates different resource sharing applications such as Ipernity, Flickr, Youtube, Vimeo, Ted, etc. and provides features for organizing and sharing these resources in a collaborative environment. When a user issues a query to the system, it returns an aggregated result set comprising of resources from the different resource sharing systems mentioned above. For example, if it's an image search then resources are returned from Ipernity or Flickr, and if it's a video search then Youtube and Vimeo returns relevant resources. I implemented a REST client for this system in order to capture the search context: the issues queries; set of resources that was displayed but neither clicked nor saved; the resources clicked or saved by the user; for how long a resource was viewed; as well as the annotations for a search using comments or tags. All this data related to the search context is then wrapped in a format acceptable by the RESTful web service and sent for the purpose of storing and managing this data. The web service also provides endpoints to support visualization of this data, which will help users, rebuild context of a search as well as help users analyze if the resources or results considered useful initially was indeed the best available resources for a particular query or if they should explore further.

## 1.2 Motivation

User's information needs that are too complex that it cannot be answered by a single query or singe page of search results can be referred to as research missions. Kellar *et al.* in [5] described information gathering on the web as a collection of information from multiple sources to satisfy a particular information need or task. In the user study conducted over a period of one week of 21 university students in 2005, it was reported that information gathering accounted for about 13.4% of overall web usage and was the fourth most important activity after transactions (46.7%), just browsing (19.9%) and fact finding (18.3%). We can understand that research missions can occur only during search sessions, and not during overall web usage. When authors manually analyzed query sessions over a period of 3 days they observed that around 10% of search sessions were research missions and around 25% of query volume is posted in these sessions [6]. This highlights the need for a search history based graphical user interfaces to support research missions.

Research conducted on task interruptions [7, 8] shows that users frequently switch tasks or tasks are interrupted by some external circumstances. There is usually a long delay before a task is resumed; even interleaving tasks leads to the change in state of the web browser. So it is important for the user to remember the current step of the task which includes what all queries were issued and the results that were considered relevant while returning back to a task after a delay. The problems that could arise during an information gathering task resumption is the duplication of search queries, this is observed from the user study conducted by Kellar *et al* [5] that 58.8% of information gathering sessions comprised of tasks that were repeated at least once. Web log analysis of Yahoo! Search engine of 114 users over a course of a year and user surveys showed that users frequently re-enter previously issued queries in order to re-find information, which could be helpful to resume a disrupted task [9, 10]. Thus a search history comprising of previously posted queries along with the resources returned could help in re-finding of information and task resumption during the search.

Komlodi carried out a case study of search histories and how it could be used for task management support in information seeking tasks [11]. The findings of this study reflects on how search histories can support complex search tasks by providing support for the planning of actions in order to monitor progress, integration of tasks and recreation of context for interrupted tasks. It also highlights the need for note-taking and annotation tools integrated into search histories to help users record their interpretations of a particular search and the result sets returned. There also needs to be mechanism for the user to review and interpret results from queries, judging the relevance of the resources or information found which will help the user evaluate if the path on which they are searching is clear or if they have already found the information they were looking for. This study shows the need for a system that will allow commenting and tagging of search histories for task representations, and re-visitation of a past result set in order to verify the search already carried out as well as to monitor the progress of a search task.

In social science research while building a multimedia corpus it is important to keep track of the search context and provide means to revisit this saved search context later. While studying the built corpus, it is crucial to understand the set of resources from which a particular image was chosen for the corpus. Thus the user should be able to revisit a result set corresponding to an issued query. It is also relevant to observe how the set of resources have changed over time by comparing the similar queries from the search history by highlighting the new resources which were missing from the previous set. There should be a mechanism for the researcher to document his thought processes during a particular search, queries or particular findings by giving comments or tags. Furthermore researchers should be able to share the set of resources from which a particular image was chosen with fellow researchers.

## 1.3 Problem Description

These days' keyword search applications have simple logging mechanisms to capture search context such as the issued queries and the resources that were clicked or considered relevant. But this search context information is not sufficient to understand from which set of resources a particular result or results were chosen for building a multimedia corpus. LearnWeb2.0 system which is considered for our scenario has a simple logging mechanism that keeps track of the queries issued, which resource was saved and to which group, various actions on groups such as group creation, deletion, and the management of resources within groups. This logging feature of LearnWeb is basic and provides more focus on actions performed within the system with less focus on actions performed during a search. The users don't have access to view these logged actions, it's only the admin who can view it by accessing the MySQL database. While searching the user doesn't have a convenient way to access the search history, and if the user needs to access it, he has to open the history tool of the browser which has history entries of browsing interleaved with the search links for the LearnWeb2.0 system which is challenging to find. If the user wants to view the information of which resources were clicked and for how long those resources were viewed as part of a previous search, it is not possible from the search history of the browser and the existing logging functionality.

Thus there was a need to integrate modules which would focus on capturing events which took place during the search for the LearnWeb2.0 system. Additionally search history graphical user interfaces (GUI) should be integrated into LearnWeb2.0 system to assist the user in viewing the information that is captured part of the search while the user is searching or is navigating through the system. There was also a lack of information in the search history of the browser as to why a particular search was carried out by the user. Therefore the user should be able to add more information or thoughts about the search using comments or tags.

## 1.4 Related Work

Prior approaches to record or manage information seeking tasks while searching or browsing the web could be classified into three main categories:

***Link centric*** approach helped users store relevant links while browsing the web or performing a search. Bookmarks provides support for this approach and it could be used to store URLs of potentially useful results as well as queries while searching over a period of time. Few drawbacks of using bookmarks for the scenario of keeping track of a search, is that, to remember a query it would be required to bookmark the search results page, and as the user keeps marking search results pages and useful links from different queries they become interleaved and makes it difficult for the user to understand. One way to tackle this problem is to create folders for an information need and then add the results accordingly to the folders but this requires too much effort from the user. There has been research where they have found ways to automatically structure bookmarks [12, 13], but it has been shown that it becomes harder to manage these bookmarks as the number reaches over a few dozen [16]. Google Notebook [14] was a browser plug-in launched in 2006 that allowed users to store URLs in a collection of notebooks, and they could annotate it with comments and labels. From the following we can see that link centric isn't the best approach in terms of usability to keep track of search tasks.

***Page centric*** approach allowed users to annotate and highlight paragraphs or passages in web pages in addition to saving the interesting results; this provided the user with better understanding of why a particular result was useful. An application which mimics this approach is Diigo [15], it allows users to highlight, comment and add sticky notes to parts of the web page which persists in the user's library. Hunter Gatherer [17] is another such tool which allows users to accumulate components from within web pages into a single page which contains links back to the original sources. All these tools help the user keep track of the information gathered which works as a good information management tool, but it fails to link back to the search process that leads to the discovery of this information and thereby not assisting in keeping track of the steps taken during the search.

**Search centric** approach is specific to search results that helps keep note of the search context: the queries posted and the useful results corresponding to those queries. In the previous approaches the information is obtained from arbitrary pages which could be reached by any means not necessarily from the search results. The first attempt in this direction was *SearchPad* [18] which assisted users in keeping track of their search progress, by storing the queries issued by the user along with the result pages the user visited or liked in the context of each query. In addition to this it stores the time spent viewing the result, provides a view to display this stored information so that a user could regain context of an earlier search process as well as reissue a query in order to explore further and finally a way to edit this stored information. The interface provided by the $S^3$ system [19] is different from *SearchPad* as it also records the results retrieved for each query along with the web pages visited by the user, additionally the user can also comment on a particular search result. Distinctive feature of this system was that the user could reissue a query from the detailed representation of the history and check if there are any results that were not among the top ten results when the query was initially executed, plus users could share their stored investigations with others via email, etc. Donato *et al.* [20] developed *Yahoo SearchPad*, a system which automatically identifies search tasks and prompts the user to take notes with a workspace already populated with queries and results visited that is related to that task. This system uses topical coherence between consecutive queries to automatically segment search tasks and provide the user with a new search pad. *SearchBar* developed by Morris *et al.* [21] proactively stores the queries issued, the corresponding URLs visited and the ratings of these links in a hierarchical structure. Users can create a topic, add notes to the topic as well as edit the search history by deleting topics, or rearranging the queries and URLs among different queries.

The search tracker service that I built for LearnWeb2.0 system incorporates various features of these existing systems and provides some unique additional functionality. The search tracker records search context features similar to what is stored by *SearchPad* [18] and also provides annotation functionality like the $S^3$ system. In the $S^3$ system, reissuing a query compares the current top ten results with the top ten results returned when the query was issued earlier to check which the new resources were. But with the search tracker service we can compare the first 'n' current resources returned for a query with the 'n' resources returned for a similar query posted in the past. In addition, to this the search tracker service also allows users to enter tags for queries, which is similar to the topic creation functionality of *SearchBar* but the tags are carried forward to the next query thereby creating search task trail, which can be edited by the user if a particular query belongs to a different task. Search tracker uses the user management functionality of the LearnWeb2.0 system to provide online sharing of search contexts for further collaboration or verification of collection building processes.

## 1.5 Systems Requirements

### 1.5.1 Software Requirements

- Eclipse IDE for Java EE (standard 4.3.2)
- Mojarra JavaServer Faces (JSF2.0)
- PrimeFaces (component library for JSF2.0)
- Java SE 7
- Eclipse Subversive plug-in and svn connector
- Java API for RESTful services (JAX-RS)
- Jersey framework for RESTful client
- JQuery 2.1.1
- MySQL server 5.0.95
- phpMyAdmin 4.2.0
- Apache Tomcat server 7.0.54
- Any web browser ( Chrome, Firefox, Internet Explorer )

### 1.5.2 Hardware Requirements

- Minimum 2GB RAM for eclipse IDE
- Any modern processor with clock rate 1 GHz or higher
- Minimum hard disk space required 700 MB
- Display properties 24-bit color depth
- Video adapter minimum 64MB RAM;

## 1.6 Report Organization

This report focuses on the detailed implementation of a search tracker service that could be integrated with any system such as the LeanWeb2.0 platform. There is also information of the various user interface tools implemented into the LearnWeb2.0 system in order to assist the user in revisiting the information captured by the search tracker service. The overall objective of this effort is to plan and implement tools that assist the user in keeping track as well as recreating context of a previous search process.

The current chapter deals with the introduction of the system, motivation, problem description, related work, software and hardware requirements for the application. Accordingly, the balance of this report has been organized into:

The Second Chapter: Overview of the proposed work contains the problem description and its related concepts along with the architectural design of the proposed system.

The Third Chapter: Analysis and Design explains the detailed design of the system along with the requirement analysis of the system and module description.

The Fourth Chapter: Implementation, explains how the project is implemented along with unit test cases.

The Fifth Chapter: Results and Discussion discusses the affectability of the proposed system.

The Sixth Chapter: Conclusion and Future Enhancements speaks of further work that can be carried out in enhancing the system which could provide more assistance to the user.

# CHAPTER 2

# OVERVIEW OF THE PROPOSED WORK

## 2.1 Introduction of problem and its related concepts

Today, keyword search applications have simple logging mechanisms to capture the actions of the user while searching, which we refer to as search context. The search context data captured comprises of the issued queries and the result links clicked by the user. This information is not sufficient for social science research, where they have to understand the steps taken in order to build a multimedia corpus plus a way to review how this corpus was built. Thus taking this scenario into consideration, the search context data that needs to be captured are: the queries issued, the resources returned and displayed for that query, the resources clicked or saved, for how long a resource was viewed and finally user annotations such as comments or tags. To help capture this search context, a RESTful (Representational State Transfer) web service was built so that we could expose the services of search tracking as a REST API which could be used by keyword search systems like LearnWeb2.0, the only requirement was that the systems have to implement REST clients that help communicate between the system and the search tracker service.

Representational State Transfer (REST) is motivated by the way the web is designed, that is, there is a network of web pages (resources) and the user progresses through an application by selecting links (states of resources), resulting in the next page being transferred (new state) to the user and rendered. REST was considered over SOAP and WSDL style web services because it is simpler to use and is based on the resource oriented model, which consists of resource states and transfer of these states using HTTP methods to various clients that are implemented in different languages. Search Tracker is built in Java using the JAX-RS framework, but it could be accessed by any system developed in different languages apart from Java, which was the motivation behind providing search tracker as a web service as it could be seamlessly integrated into any

keyword search system. Other key characteristics of RESTful web services is the use of HTTP methods in a one-to-one mapping with create, read, update and delete (CRUD) operations on a resource or a batch of resources. It is also stateless, that is, each request from the client to server should include all the information needed to understand the request and no context information is stored at the server, the advantage of this is better performance and more scalability.

The next problem was the transfer of information or data from the LearnWeb2.0 system and the search tracker web service. For this purpose a LearnWeb2.0 REST client was built to wrap the data in a format acceptable by the search tracker service (XML or JSON) and is communicated to the service. The LearnWeb2.0 system is implemented in JSF2.0, a Java specification for building component based user interfaces for web applications. Thus the RESTful client for LearnWeb2.0 is built using the Jersey framework in Java. Jersey framework is helpful in building REST clients as it provides support for JAX-RS APIs and serves as a JAX-RS reference implementation.

Search history modules to access the search contexts for a particular user that was captured using the search tracker service had to be integrated into the LearnWeb2.0 system. LearnWeb2.0 is developed using JavaServer Faces 2.0 (JSF2.0), a Java UI component framework for building dynamic pages for web applications. It provides an API for creating; managing and handling UI components and tags which help build components for a web page. PrimeFaces is a light weight open source component suite for JSF2.0 which is used to build rich set of components for LearnWeb2.0 application. It is a lightweight, one jar and requires no dependencies; it also provides in-built AJAX based upon the JSF Ajax APIs and a skinning framework to allow the developer to design the visual theme. Thus the search history modules for displaying the search contexts is developed using JSF2.0 and PrimeFaces. The search results page interface is modified to display search context information such as the complete search history, queries similar to the one posted and information of the user interactions with the current search results such as the resources clicked, saved and viewing time of the resources. The user could also provide additional information about the current search through tags and comments provided part of the interface. Display of the similar queries features provides the user with the option of comparing the resources of a previous query in the search history with the current resources being returned, giving the user a means to gain perspective of how the result sets have evolved over time. The next challenge was to provide the user with a view to explore the complete history with the possibility of editing the history to remove search contexts no longer relevant to the user. So an additional interface was designed in order to provide this functionality to the user along with links for each query, which redirects to an interface similar to the search page that displays the resources that were returned corresponding to that query along with other details of search context and an opportunity to carry out further investigation.

## 2.2 Overview of the proposed system

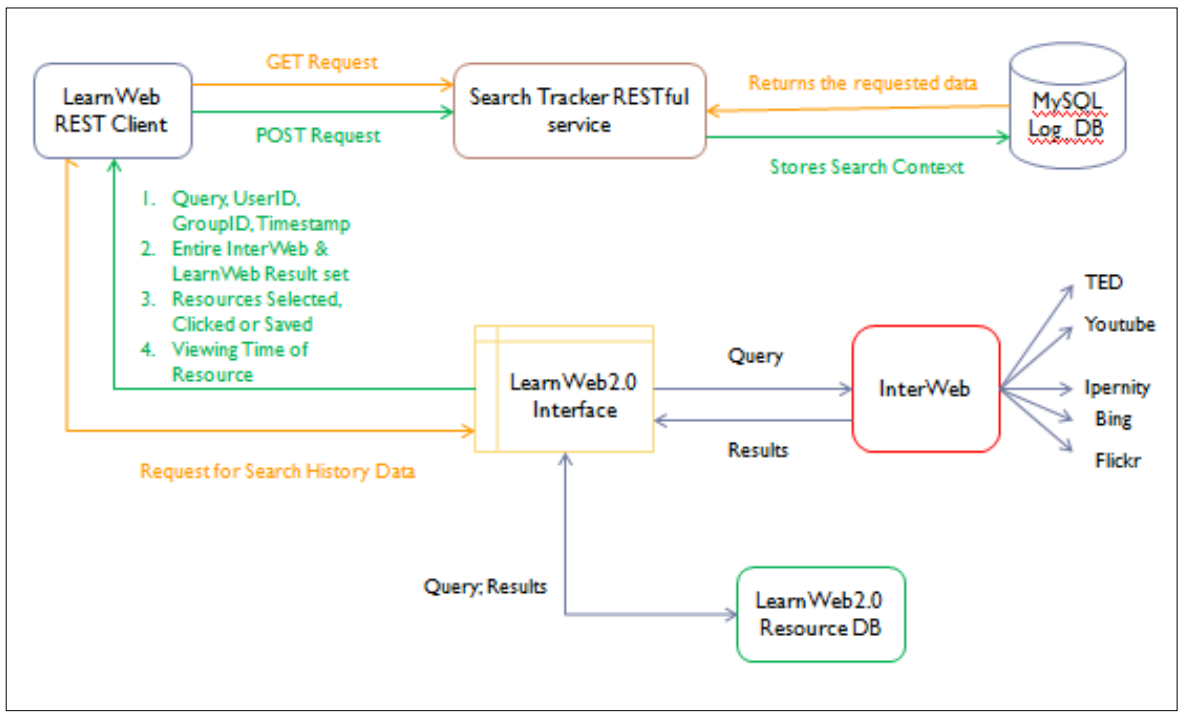

**Figure 2.1: Architectural design of the system**

The above diagram shows how the LearnWeb2.0 system interacts with the search tracker RESTful web service. Initially the query posted by the user, along with the user id, group id and the timestamp is passed to the LearnWeb2.0 REST client that sends a POST request to the search tracker service in the form of XML/ JSON. The search tracker receives this XML/ JSON data and creates a database connection using JDBC and stores the query data in the respective table. The query posted by the user is sent to the LearnWeb repository of resources as well as InterWeb which is a Web 2.0 service adapter providing a seamless interface for accessing Web 2.0 data from applications such as Youtube, Flickr, Ipernity, Vimeo, Bing and TED. The resources returned by both the LearnWeb repository and InterWeb provides an aggregated result set which is forwarded to the REST client. This result set is temporarily stored until a query change occurs or a timeout of 10 minutes passes after which it is sent as a batch to the search tracker through a POST request which is then stored into the respective table by the service. The resource click and the viewing time are stored temporarily in the REST client and posted to the web service after a request to post the result set occurs. In context of LearnWeb2.0 we store the resource save event which captures when a particular resource is saved to a group or user's personal resources this helps provide additional information for the search context highlighting the relevance of the particular resource to the user, this is also recorded in a similar manner to the click event.

In order to display the complete search history in the user interface of the search page, LearnWeb2.0 sends a request to fetch the data to the REST client which in turn fires a GET request to fetch the complete search history corresponding to that particular user. The search tracker service then retrieves the respective records from the database and returns the data back to LearnWeb2.0 system. The request for retrieving the similar queries and the data for exploring of the search history is similar to the previous one. When the user wants to revisit a previous result set, the result set id of the corresponding query is sent to the LearnWeb client in turn firing a GET request with these parameters. The search tracker service returns the resources corresponding to that particular result set id, and this is forwarded back to LearnWeb2.0 system which displays it using the resource display templates of the search page. Users are allowed to further analyze and annotate this result set as it being viewed, which leads to subsequent POST requests from the REST client with the current data containing the new batch of resources clicked or saved and new comments or tags added or removed to the already existing annotations for that search context. All the data communicated between the LearnWeb REST client and search tracker service is either in the form of XML, JSON or plain text.

# CHAPTER 3

# ANALYSIS AND DESIGN

## 3.1 Brief Introduction

The design and the workflow are vital to the smooth working of any application. The design architecture throws light upon the different modules that need to be developed and how the modules depend on each other, thus providing the developers with the right direction to develop a successful application.

The initial analysis carried out highlighted the importance of a well documented search process which helps a user understand how a corpus was built as well as review the built corpus. A detailed history of the search activities should be displayed which includes not only the query posted by the user but also the resources clicked or saved and the viewing time of a resource. In order to keep track of how and why a corpus was built, the user should be provided options to document their thoughts and decisions regarding the search and particular findings in the form of comments or tags. There also needs to be ways to edit the search history in order to remove or delete search contexts or queries, so that only the search contexts which are relevant to the user are presented or displayed. While searching, it is important to provide a feature to compare the current query with similar queries in history in order to highlight those resources that are new in the current search results, this helps build perspective of how a corpus has changed over time. It is possible a user would like to revisit a previous result set in the search history in order to carry out further investigation and select more resources relevant to that search, or annotate it differently.

The rest of the chapter comprises of requirement analysis which is needed to build a successful working application that meets the various needs of the user and performs in a suitable manner. The design focuses on the architectural styles of the various components used to build the search tracker service and the search history modules for the LearnWeb2.0 system. Finally module description highlights the functionality of the different modules integrated into the system.

## 3.2 Requirement Analysis

Requirement Analysis in software or systems engineering comprises of those tasks that helps in determining the needs and constraints that is required for a new or modified product. It needs to take into account the various conflicting requirements of various stakeholders that are persons or organizations with an interest in the application. This step requires documentation of the requirements, analysis of these requirements to see if they are clear, unambiguous and achievable.

### 3.2.1 Functional Requirements

These requirements generally comprise of the functions that needs to be provided by the application that is developed. This basically provides the description of the required behavior of the application.

- The system should be able to record the search context for a particular search activity.
- The system should provide a suitable interface to view the complete search history for a particular user.
- A user should only be able to view their search history and not any other user's.
- User should be able to edit the search history, that is should have the possibility to delete search processes or queries.
- Users should only be able to view those result sets shared with them.
- It should not be possible for a user to view a result set belonging to the search history of another user unless it was shared with them.
- Only users who want to view the search history should be provided with the functionality.
- The shared result sets should reflect the web investigations carried out by other users on the result set as well as the added comments or modifications in tags.
- The functionality provided by the search results page should also be consistently reflected in the view of the previous result set pages.

### 3.2.2 Non-Functional Requirements

These requirements provide constraints which can be used to evaluate the operation of a system or application. It elaborates upon the performance characteristic of the system.

- The system should perform functions in real-time.
- The system should be scalable.
- The system should a good response time.
- The usability of the system should be intuitive to the user.
- The system should be reliable.
- All the data logged by the search tracker should be displayed to the user consistently.

### 3.3 Detailed design of the system

The entire system for logging the search process and the user interface tools to visualize or display this information was developed following the various steps in the software development cycle.

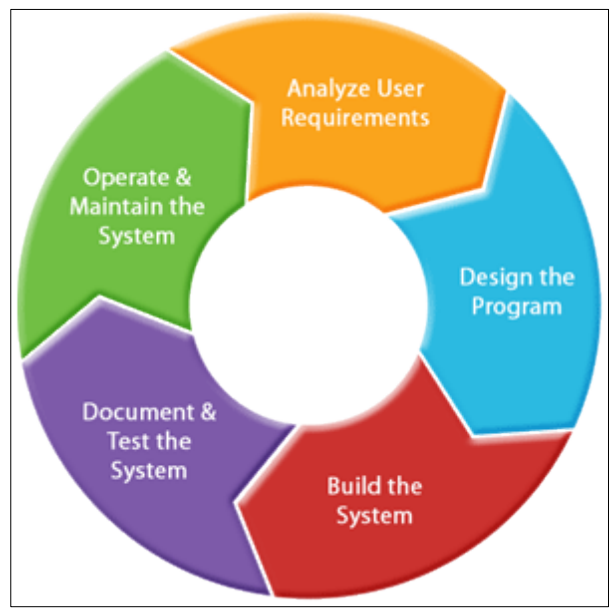

**Figure 3.1: Software Development Life Cycle**

Initially the user requirements for the purpose of tracking search history was obtained in discussions with social scientists who extensively analyze their search processes and this provided a concrete base to develop a system to help assist the user in keeping track of search processes and further analyze the result sets obtained during previous search tasks. Then the architectural design of the system (Figure 2.1) was created which gave additional insights into the different modules required by the system and in what sequence the modules needed to be implemented. Then the system was developed, first the search tracker service was implemented and then the LearnWeb2.0 client along with the search history modules to display the search contexts was developed in parallel.

### 3.3.1 REST Architectural style

A web service is a system which supports interoperability between machines over a network. It provides an interface described in machine process able format, and messages are conveyed using HTTP standards. REST-complaint web services are a major class of web services, which manipulates representations of web resources using a set of stateless operations. REST describes software architectural style in which web services are designed with a focuses on system resources (data and functionality), including how they are addressed using Uniform Resource Identifiers (URI's) and transferred over HTTP using REST clients that are implemented using different languages. It provides uniform interface semantics such as create, retrieve, update and delete rather than application specific interfaces and manipulates resources by the exchange of representations.

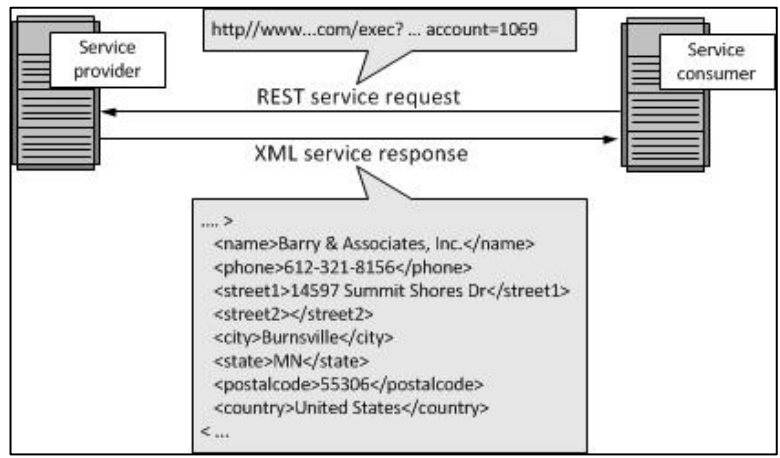

**Figure 3.2: REST – complaint Web Service**

The various constraints applied to the REST architectural style are as follows:

- *Uniform Interface:* This defines the interface between the client and the server. It helps simplify and decouple the architecture allowing each component to evolve independently. Individual resources are identified through URI's, but the resources themselves are separate from the representations sent to the client. Resources are manipulated using a fixed set of simple operations PUT, GET, POST and DELETE providing functionality similar to the HTTP methods.

- *Stateless:* The communication between client and server does not depend on the client context being stored on the server between requests. Each request from the client has enough information needed to process that request on the server.

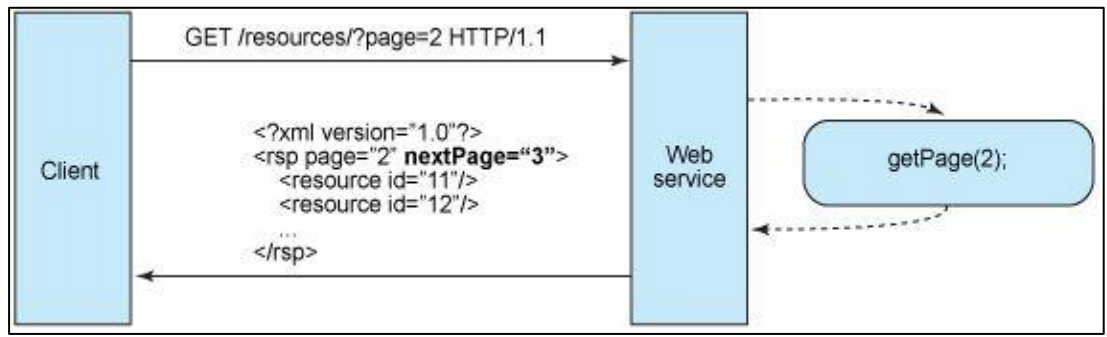

**Figure 3.3: Stateless design of RESTful web services**

- *Client – Server:* A uniform interface separates the client and server. This indicates that clients need not be concerned by data storage, which increases portability of client code. The server need not be concerned by the user state or interface, thus making the implementation of the server simple and thereby more scalable.

- *Cacheable:* Client responses can be cached. Thus it is important for responses, to indicate if a response is cacheable, in order to prevent a state being used by the client in future requests.

- *Layered system:* Client need not be directly to the connected to the end server. There can be multiple intermediate servers in between which helps increase scalability.

- *Code on demand:* Servers can customize the functionality of a client by transferring executable code.

All these constraints are taken into consideration while developing the search tracker service.

### 3.3.2 LearnWeb2.0 System Architecture

LearnWeb2.0 is a Rich Internet Application (RIA), which provides a nice interactive user experience much like the features and functionality provided by desktop applications. It provides a rich experience through a light weight web browser without the requirement of software installation on the client side. One prominent feature which makes RIA different from other web based applications is that it acts as a client engine between user requests and an application server. LearnWeb2.0 is modeled using the MVC paradigm.

Model – View – Controller (MVC) is a software architectural pattern, considered as an architectural design to implement user interfaces in applications. It separates modeling of the domain (application logic for the user) from the presentation (user interface), thus enabling the development, maintenance and testing of each module independently.

- *Model:* The model manages the knowledge, that is, the behavior and data of the application domain. The model receives requests for information state from the view and requests to change the state of information from the controller. In event-driven systems, the model updates the view if there is a change in information state so that the users could react.
- *View:* It requests information from the model and provides an appropriate representation. There is a possibility to hide certain features and highlight important attributes, thus acting as a presentation filter. Multiple views could exist for the same model, thereby providing different user interfaces according to the requirements of the application. For example, the user interface for the complete search history in the search history page and the interface for the option to explore the search history depend upon a common model.
- *Controller:* Is the link between the user and the LearnWeb2.0 system. All the user actions are received by the controller and it is forwarded to the model to update its state (in case of document editing), or could also be sent to the view (in case of document scrolling).

The MVC framework used for implementing the LearnWeb2.0 interface is JavaServer Faces 2.0 (JSF2.0).

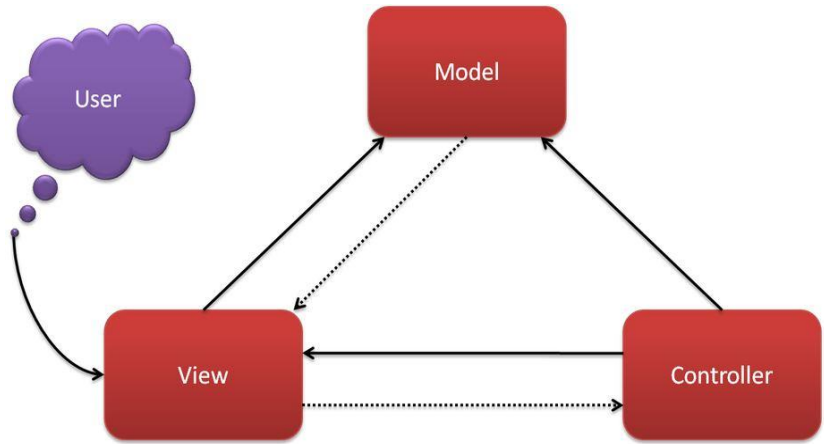

**Figure 3.4: MVC Architecture**

The search history interface tools integrated into LearnWeb2.0 are designed in accordance with the MVC paradigm of implementing user interfaces.

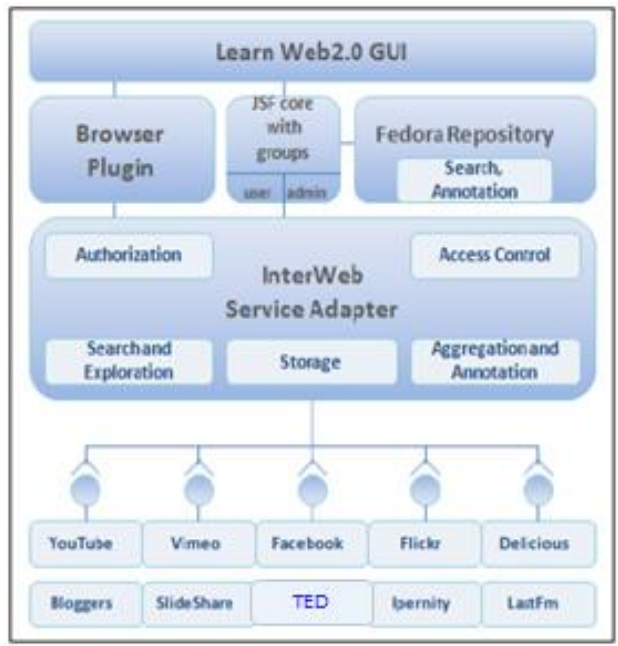

**Figure 3.5: LearnWeb2.0 System Architecture**

The LearnWeb2.0 user interface provides functionality such as searching for resources, organizing resources and sharing resources. The Web2.0 service adapter InterWeb is used to provide a seamless interface for accessing resources from Youtube, Flickr, Ipernity and many more sources. The fedora repository stores all the resources saved as part of LearnWeb2.0.

## 3.4 Sequence Diagrams for the proposed system

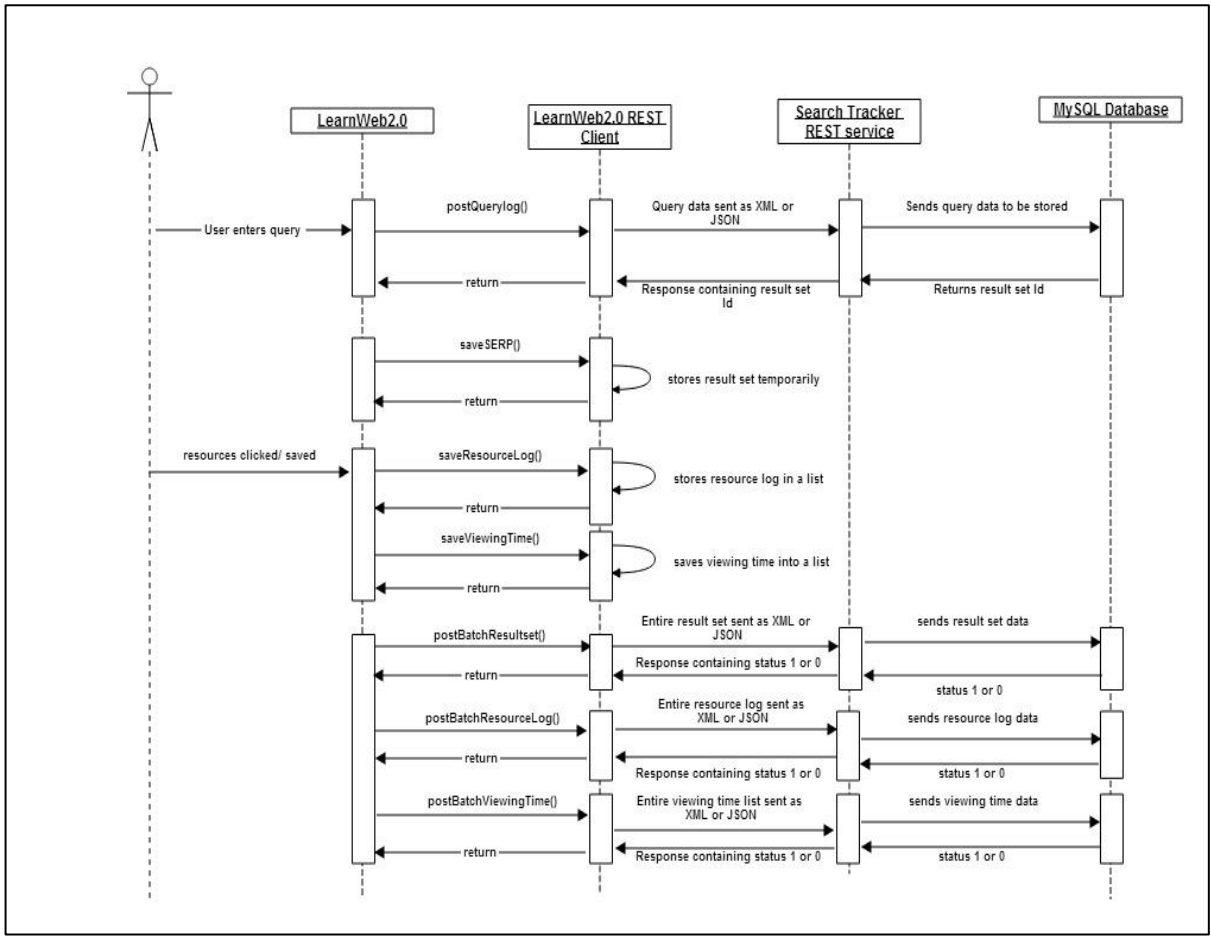

**Figure 3.6: Sequence diagram for search log process**

This above sequence diagram highlights the sequence of events that take place while the user is searching. Initially the query posted by the user is sent to the REST client where the data is wrapped into an acceptable format and sends it to the search tracker service which stores this data into the MySQL database. The aggregated result set returned corresponding to that query is saved temporarily at the client, along with the various user interactions on the resources such as resource clicked, saved as well as viewing time. After a particular timeout, query change or session end, the entire result set is sent as a batch to the service which stores it in the database. Similarly the resource click and save events is batched as a resource log list and the viewing time as well which is then sent to the search tracker service that stores this information in to the respective tables in the database.

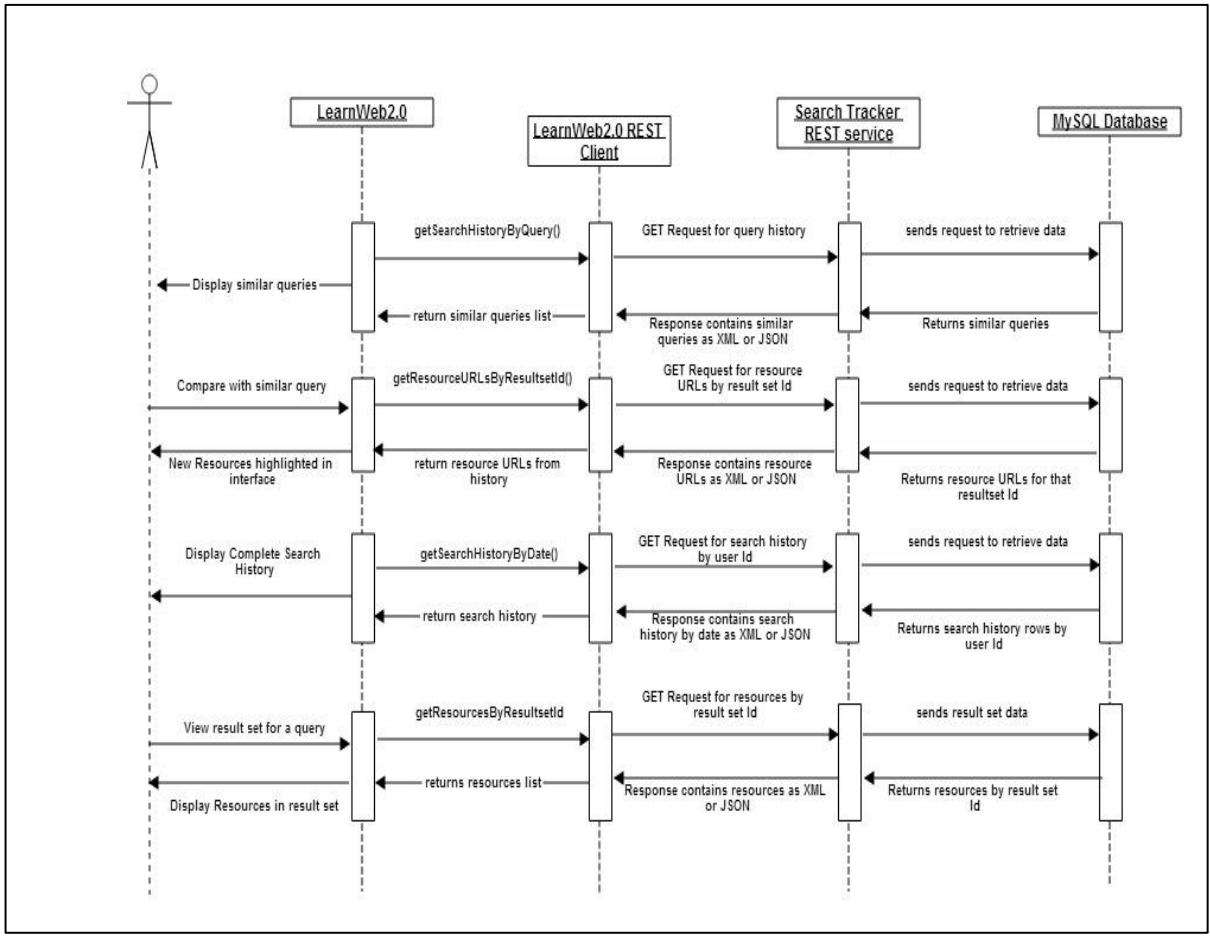

**Figure 3.7: Sequence diagram for displaying search history information**

This sequence diagram brings to focus the different steps involved in displaying the search history for that user. Initially the similar queries is displayed when LearnWeb2.0 calls the corresponding method in the REST client that in turn fires a GET request to the search tracker in order to retrieve the corresponding similar queries from the database, which is then sent in the form of XML or JSON back to the client and finally the list of similar queries are sent to LearnWeb2.0 which displays it to the user. The user could compare the result set of a similar query the current resources, this requests a set of resource URLs corresponding to that result set id from search tracker which extracts it from the database and forwards it to the client, which then transfers it to LearnWeb2.0 thereby highlighting any new resources in the current set. In a similar manner the complete history for a particular user and the result set corresponding to a query in the history is retrieved from the service and sent back to LearnWeb2.0 to be displayed.

## 3.5 Use Case Diagrams

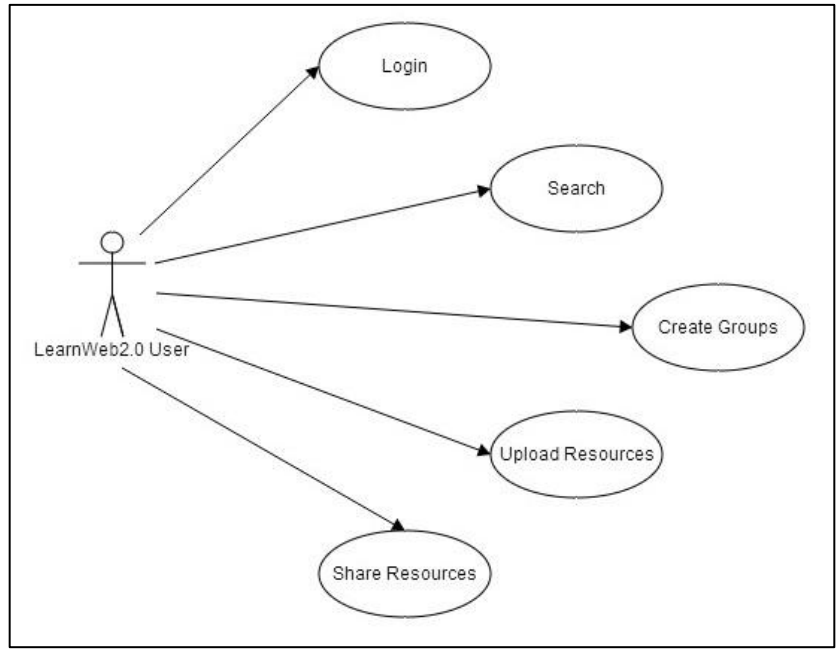

**Figure 3.8: Normal LearnWeb2.0 User**

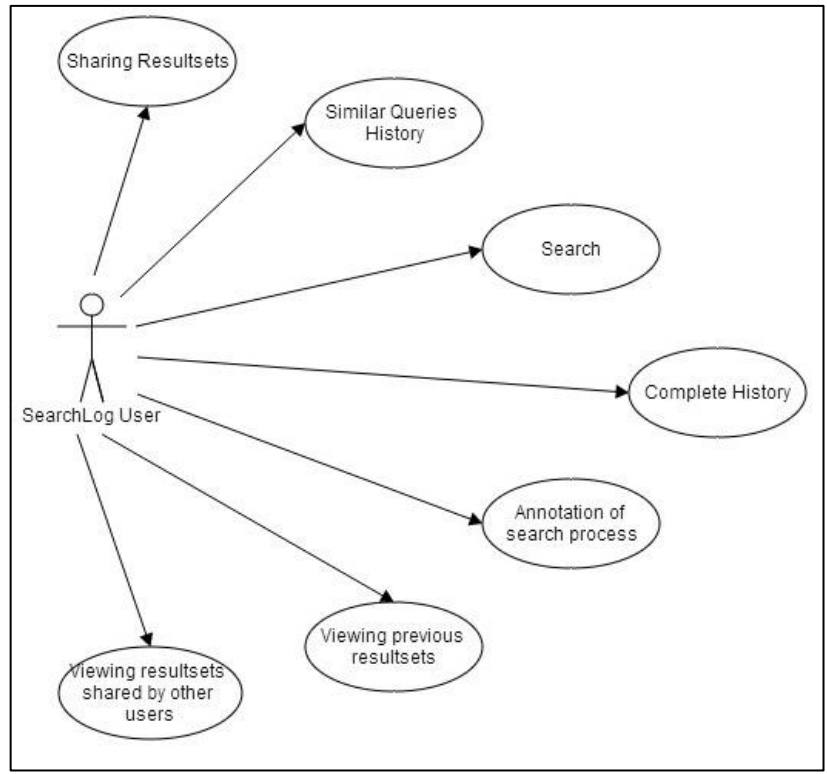

**Figure 3.9: LearnWeb2.0 user with search tracker functionality**

### 3.6 Module Description

The proposed system to log search processes and to provide further analysis of this search history data is mainly comprised of three modules: the search tracker REST service, LearnWeb2.0 REST client and the user interfaces for displaying of the search context data collected during the search process.

1. *Search tracker REST service*

The search tracker service should be designed keeping in mind the principles of REST Web service design which are as follows:

- First step is to identify all the conceptual entities which need to be exposed as services.
- Each service or endpoint identified should have a URL associated with it, so that it could be accessed by a client.
- Resources need to be categorized according to whether clients should just receive a representation of a resource or if the client could actually modify (add, update or delete) a resource. For the former category, these resources are made accessible by HTTP GET, where as the latter is made accessible using HTTP PUT, POST and DELETE.
- All resources accessible via HTTP GET method should be side effect free. This means that invoking a resource should just return a representation of the resource, it should not result in modification of the resource.

The search tracker service thus exposes various endpoints which help in initially storing the search process of the user, that is, the query posted by the user along with the results returned, the resources clicked or saved and for how long they were viewed. It also helps store the various annotations by the user in the form of comments or tags for a search process. Once all the information is stored, the service provides endpoints to make this information of search context available to the user. These endpoints are exposed using a set of URLs which could be accessed by any client. For retrieving the information GET requests are implemented where as to post the search information PUT or POST requests are used.

*2. LearnWeb2.0 REST client*

The LearnWeb2.0 system communicates with the search tracker service using the LearnWeb2.0 REST client. The various search information captured within the LearnWeb system is forwarded to the client which then represents this data in either XML or JSON format and forwards it to the search tracker service for storing. The search tracker then sends responses back to the client indicating if a particular request to store the information was indeed successfully stored or not. In certain cases, the client stores the search context information temporarily before sending this data as a batch to the search tracker service. This is done in order to manage the overhead in communication between the client and service, as well as to reduce the load on the search tracker service. When the LearnWeb2.0 system requests search history information from the client, the client sends the appropriate GET request to the search tracker service. The search tracker service retrieves the appropriate data from the database and sends this data back to the REST client using XML or JSON format. The REST client once it receives the data it is stored in various data structures such as hash maps or lists which are accessible to the LearnWeb2.0 system.

*3. Search history modules for LearnWeb2.0*

The search history modules integrated into LearnWeb2.0 provides or assists the user in viewing of the search history information. Initially the search history modules were designed for the search results page where the user enters a query to start the search process. When the user starts the search process the search history information is hidden to the user, but if the user does want to see this information, there is a search history tools button which on clicking displays this information alongside the search results as a right panel.

This right bar or panel provides a tab view consisting of various tabs such as similar queries, complete history, and current search. The first tab similar queries, displays the query history of queries that are similar to the query that is posted currently by the user. The user could click on any one of these similar queries to compare the resources returned and viewed by the user for that particular query with an equivalent number of first top resources that is returned for the current query posted by the user. The new resources in the current set of results which were not present in the previous result set are highlighted; this helps the user gauge how much the corpus has changed for a similar query over time. The complete history tab displays the complete history of the user along with the resources that were clicked or saved. By clicking on any one of these results the user will be redirected to a new page which displays the resources that were returned for that query in a layout similar to the search page. This provides the user with the opportunity to understand the previous search conducted as well as gives the user a chance to analyze this result set further. The current search tab provides the user with the information of his current search activities showing the various resources that were clicked and for how long they were viewed along with the resources that were saved. In this tab the user could annotate the search by providing both comments and tags.

The next interface is the "explore search history" interface, which is independent from the search page. This interface displays the complete history of the user implemented as a paging mechanism. Each entry of the search history displays the query posted along with the resources that were clicked or saved and the corresponding time at which these events were recorded. This helps the user to keep track of his search task as well as understand the steps that were taken to build the multimedia corpus. There is also a mechanism to filter the search history between particular dates specified by the user, thus enabling the user to focus only on particular part of the history. Corresponding to each query in the search history, a view result set button is shown when the user hovers over that query. This button redirects the user to the "view result set" page that displays the resources that were returned and viewed by the user for that particular query.

The "view result set" interface displays the resources returned for a particular query as mentioned above. Also the various events such as the resources clicked, saved and comments are displayed as a timeline in the right panel, helping the user rebuild context. There is also an option to filter the resources displayed according to either resources clicked, not clicked or saved. There is another tab in the right panel, which allows the user to edit the tags belonging to this query as well as adding new comments to this search context. There is a share result set option in this page which provides functionality for the user to share it with others who could view it as well as provide their own annotations for this investigation. The timeline also reflects the events recorded part of the further investigation process.

# CHAPTER 4

# IMPLEMENTATION

## 4.1 Tools and Technologies Used

For implementation of the system it is of immense importance to choose the right tools to implement the different functionality provided by the system. The different tools used to implement the proposed system have been discussed below. The reason for choosing these tools has also been highlighted and explained properly.

### 4.1.1 Java Server Faces (JSF)

LearnWeb2.0 system is built using JavaServer Faces framework. JavaServer Faces 2.0 is the standard Java Enterprise Edition technology for building web user interfaces. It is a server-side component framework which helps build UI components for Java technology based web applications. It is designed to simplify the burden of developing and maintaining applications that run on a Java server and renders the UI back to the required client. JSF uses different XML files called view templates or facelets view which supports the component driven design model. LearnWeb2.0 user interface is built using a set of XML files which consists of CSS (Cascading Style Sheets) that provides the styling and JavaScript to dynamically alter the document content being displayed. The application data is stored in the MySQL database and the resource (meta-data of the resource) is saved in the fedora repository.

It includes a set of APIs for building UI components and managing their state, defining page navigation, handling input validation and events, and supporting accessibility and internationalization. JSF also includes a JavaServer pages (JSP) custom tag library for the flexibility of rendering JavaServer Faces interface within a JSP page. It provides a well defined programming model and tag libraries. The tag libraries consist of tag handlers that help render or implement the UI components in the application. It makes the building of user interfaces convenient, as the requests and response logic need not be explicitly coded.

The various advantages of using JavaServer faces are:

- Easier to construct a UI from a set of reusable UI components.
- Helps managing UI state across server requests.
- Provides a simple design to bind client side generated events to server side application code.
- Migration from application data to and from the UI is simplified.
- Custom UI components can be easily built and reused.

JSF makes it easy to reuse the UI components and enables the isolation of view and logic as it is based on the MVC web framework. Most of the complex implementation details are hidden at the view layer behind MVC 2 architecture. The degree of coupling between the UI components that represents the behavior or properties and its rendering is low. JSF runs in a Java container, which contains: Java managed beans for handling the main application specific data and logic, a custom tag library is used for rendering UI components in the view and the servlet controller provides an interface for the communication between views and the managed beans.

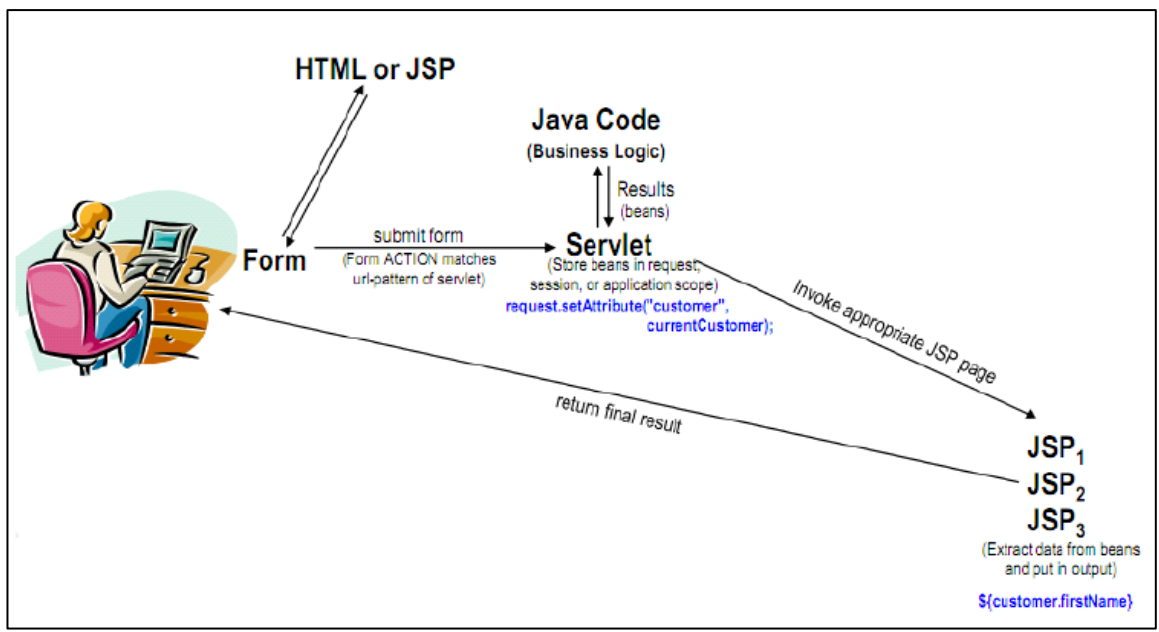

(Source: http://www.coreservlets.com/JSF-Tutorial/jsf2)

**Figure 4.1: Working of a servlet**

The JSF web application lifecycle has six phases which are as follows:

- Restore view phase
- Apply request values phase; process events
- Process validation phase; process events
- Update model values phase; process events
- Invoke application phase; process events
- Render response phase

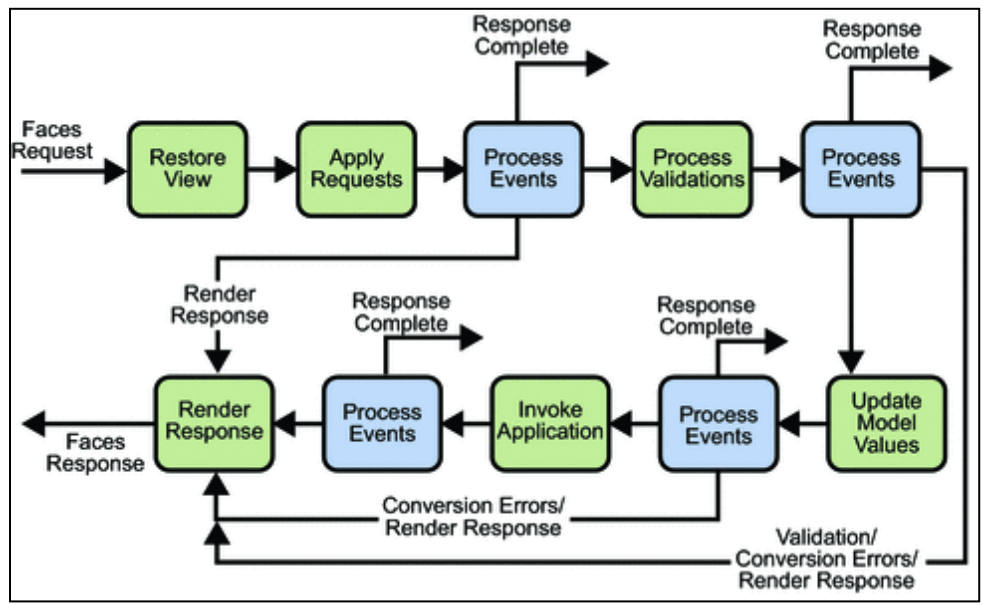

(Source: The Java EE 5 Tutorial http://docs.oracle.com/javaee/5/tutorial/doc/bnaqq.html)

**Figure 4.2: JSF Lifecycle**

### 4.1.2 PrimeFaces

PrimeFaces is the UI component suite used for building the interface. PrimeFaces is a lightweight open source component suite for Java Server Faces 2.0 featuring 100+ rich set of JSF components. The major advantage of using PrimeFaces is simplicity and performance. PrimeFaces is a lightweight library; all decisions made are based on keeping PrimeFaces as lightweight as possible. Usually adding a third-party solution could bring an overhead however this is not the case with PrimeFaces. It is just one single jar with no dependencies and nothing to configure. Integration with JQuery allows the possibility to use a lot of the JQuery libraries. It also has a rich set of skinning options, to help customize the UI.

### 4.1.3 Apache Tomcat

Apache Tomcat is a widely used web application server developed by Apache Server Foundation. This open source web server enables java code to run in, by providing pure Java HTTP web server.

The component that implements the specifications for servlet is Catalina. Catalina is an implementation of Java servlet specification. Tomcat also has some configuration files which can be used to alter the default behavior of Catalina. Its configuration and management could be handled by editing XML configuration files; Tomcat also includes tools for doing the same.

### 4.1.4 Eclipse IDE

Eclipse IDE is an integrated development environment for Java used for developing applications. It is highly extensible as it can be customized by incorporating additional plug-ins. The eclipse SDK offers IDE with a Java compiler that is incremental with the full model of Java source files. This makes analysis and refactoring of code easy. With the debug perspective provided by Eclipse IDE, it is possible to control the program execution by placing breakpoints and watching and manipulating program variables. This helps in finding out certain errors which are not visible during code review or execution.

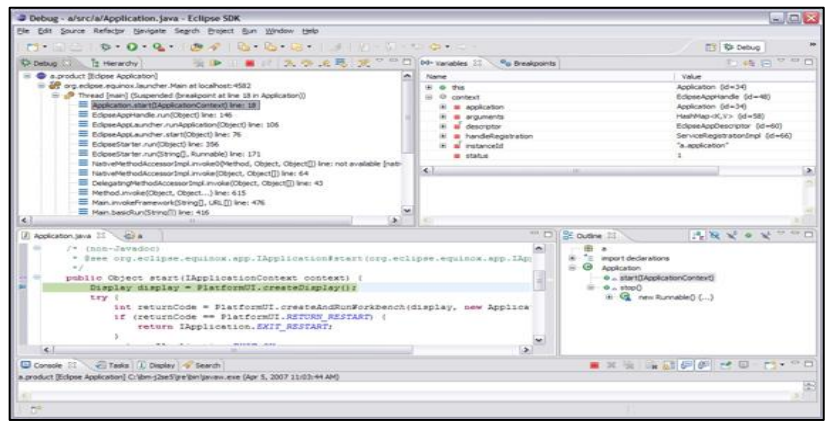

**Figure 4.3: Debug Perspective**

### 4.1.5 Java API for RESTful Services (JAX-RS)

JAX-RS implements support for annotations that are defined by JSR-311 implementation, making the development of RESTful web services using the Java programming language much simpler. In order to simplify the development and deployment of service endpoints annotations are used. These annotations along with the classes and interfaces provided by JAX-RS API, makes it easy to expose Plain Old Java Objects (POJOs) as web resources.

### 4.1.6 Jersey JAX-RS Client API

The Jersey JAX-RS client API is a Java based API used to provide access for web resources. It is not restricted to web resources particularly implemented using JAX-RS. It provides a higher-level of abstraction compared to the HTTP communication API as well as integration with the JAX-RS extension providers, in order to enable concise and efficient implementations of REST clients that leverage existing and well known client-side implementations based on HTTP communications. It encapsulates the Uniform Interface Constraint – which is a key constraint of the REST architectural style and associated data elements as Java artifacts and supports a pluggable architecture by defining multiple extension points.

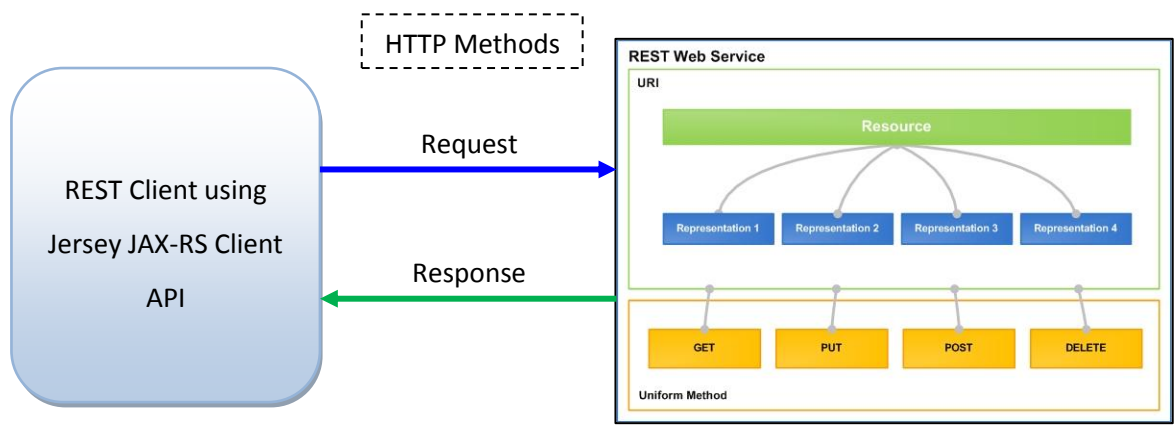

**Figure 4.4: REST Client Communicating with RESTful web service**

### 4.1.7 Java Architecture for XML Binding (JAXB)

JAXB is an XML to Java binding technology that makes the development of web services simpler by allowing the transformations between schema and Java objects and between XML documents and Java objects. XML is a common media type that is consumed and produced by the various endpoints exposed by the RESTful web services. Requests and responses can be represented by JAXB annotated Java objects, in order to serialize and deserialize XML. This JAXB objects can be used in the request entity parameters and response entities as the JAX-RS runtime environment has *MessageBodyReader* and *MessageBodyWriter* which provides implementations for reading and writing JAXB objects into entities.

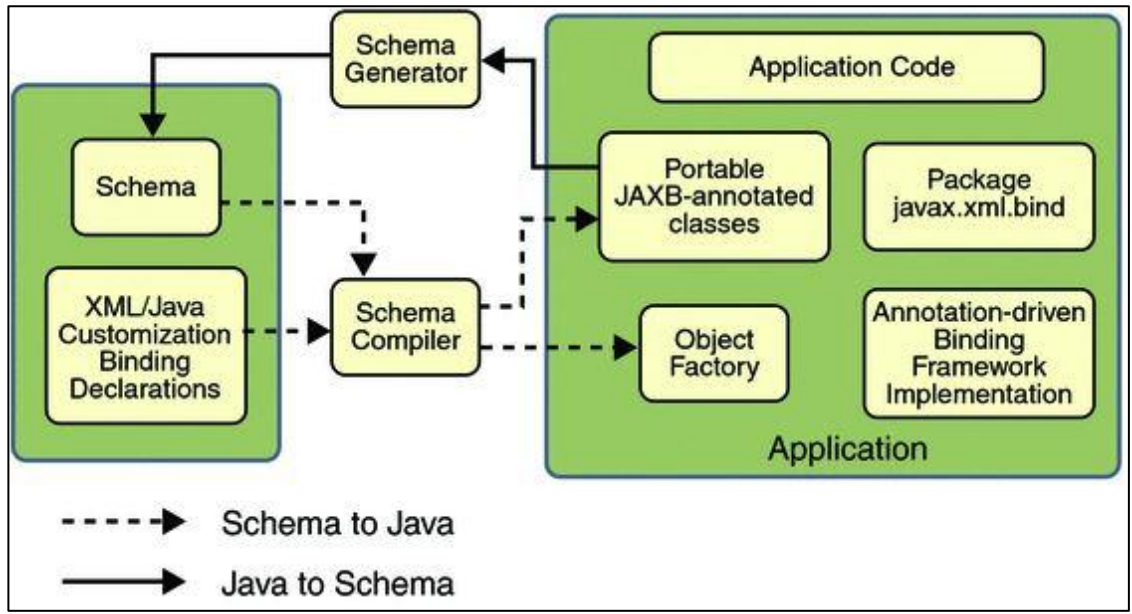

(Source: The Java EE 5 Tutorial http://docs.oracle.com/javaee/5/tutorial/doc/bnazg.html)

**Figure 4.5: JAXB Architecture Overview**

## 4.2 Implementation

The implementation of any system is very essential as the output of this stage is directly reflected in the final system. Therefore every step needs to be designed and developed keeping in mind the user requirements. The implementation of the system is divided into three main components the RESTful web service implemented using JAX-RS, the LearnWeb2.0 REST client using Jersey JAX-RS client API and finally the user interfaces for the LearnWeb2.0 system to display the search history information.

### 4.2.1 RESTful Web Service

The JAX-RS API uses annotations provided by the Java programming language to build RESTful web services. Java class files with JAX-RS annotations helps define the web resources and the actions that can be performed on these resources. These are runtime annotations which will generate helper classes and artifacts for the resources during runtime reflection.

The summary of the JAX-RS annotations that are used part of the implementation of the search tracker web service are as follows:

**Table 1: JAX-RS Annotations**

| Annotations | Description |
| --- | --- |
| @Path | This annotation is used to indicate a relative URI path where the Java class could be accessed at. Various variables could be embedded in the URI path template. |
| @GET | A request method designator assigned to a Java method which will process HTTP GET requests. |
| @POST | The Java method annotated with this request method designator |
| | will process HTTP POST requests. |
| @*DELETE* | The Java method annotated with this request method designator will process HTTP DELETE requests |
| @*PathParam* | This annotation can be used to extract a parameter for use with the resource class. URI path parameters are extracted from the request URI, and they correspond to the path template variable names. |
| @*QueryParam* | This annotation helps in extracting parameters from the request URI query parameters. |
| @*Produces* | It is used to specify the MIME media types of representations of a resource that can be generated and given to the client. |
| @*Consumes* | It is used to specify the MIME media types of representations of a resource that could be accepted or consumed by the client |

```
//Root Resource exposed at searchlog path
@Path ("/searchlog")
public class SearchLog {

/**
 * Posts the user query along with search type, userId, groupId, sessionId and
time stamp
 * @param qlinstance
 * @return Response
 */
@POST
@Consumes ({MediaType.APPLICATION_XML, MediaType.APPLICATION_JSON})
@Path ("/querylog")
public Response postQuerylog(QueryLog qlinstance) {

//Code to store the user query into the MySQL database
return Response.status(200).entity(new ServiceResponse("Database Successfully
Updated", success)).build();
}
```

```
/**
* This method is used to get the complete search history for a particular user
given the userId
* @param userId
* @return - It returns the complete history wrapped inside a SearchHistoryList
object
*/
@GET
@Path ("/searchhistory/{userid}")
@Produces ({MediaType.APPLICATION_XML, MediaType.APPLICATION_JSON})
public SearchHistoryList searchHistory(@PathParam(value = "userid") int userId){

//Code to retrieve the search history for the corresponding user from MySQL
database
return searchHistoryList;
}

/**
* Deletes the user queries corresponding to the result set IDs contained in
the result set ID list
*/
@DELETE
@Consumes ({MediaType.APPLICATION_XML, MediaType.APPLICATION_JSON})
@Path ("/deleteuserqueries")
public Response deleteUserQueries(ResultSetIdList resultsetIdList){

//Code to delete the corresponding user queries from the MySQL database
return Response.status(200).entity(new ServiceResponse("Successfully removed
the required user queries from the database ",success)).build();
}

}
```

The above code snippet gives an overview of the search tracker web service with focus on three Java methods handling the different HTTP requests from a client. The next section details the annotations used part of the search tracker service.

- The *@Path* annotation is a relative URI path. It indicates that the SearchLog Java class is hosted at the URI '/searchlog'. The java methods handling the HTTP requests could be reference by adding the URI path specified for the particular method to the URI path defined for the SearchLog class. For example, we could access the end point to delete user queries by sending a request to this URL:

  http://localhost:8080/searchlog/deleteuserqueries

- The *@PathParam* annotation is used to extract parameters from the URI path templates which are URIs with variables embedded within the URI syntax. These parameters can then be used within the resource methods. In the above snippet, we can observe that the *@PathParam* extracts the *userId* from the URI path when a request is made using this path template. This *userId* is then used to extract the search history corresponding to that user.

- Within the SearchLog resource class, request method designators such as *@GET, @POST, @PUT, @DELETE* is used to map the HTTP methods to the java language methods defined within resource class during runtime. Methods annotated with the request method designators should either return a *void*, *javax.ws.rs.core.Response* object, or *JAXB* object for the response entity bodies. Both *@PUT* and *@POST* can be used to create and update resources, but *@POST* requires the application to define the semantics while as *@PUT* has well defined semantics. Thus the *@POST* annotation is used in this scenario to store data to the database. The *@GET* annotation is used whenever the application requests data to be retrieved. And finally *@DELETE* assists the application in removing or discarding data not need by the user.

- The *@Consumes* and *@Produces* annotations are used to specify the MIME media types for the representations of resources sent back and forth between the client and the server. In the above scenario the MIME media types used are application XML or JSON, these types indicate that the representations of the resources sent or received by the search tracker web service are either XML or JSON depending upon the request by the REST client. The JAXB API supplies entity providers for the mapping between representations and the associated Java types. An example of ServiceResponse Java object with JAXB annotations that is used to send the responses back to the client is shown below.

```java
@XmlRootElement
public class ServiceResponse {

private String message;
private int returnid;

//Parameterized Constructor
public ServiceResponse(String message, int returnid){
this.message = message;
this.returnid = returnid;
}

@XmlElement
public String getMessage() {
return message;
```

```
}
public void setMessage(String message) {
this.message = message;
}

@XmlElement
public int getReturnid() {
return returnid;
}

public void setReturnid(int returnid) {
this.returnid = returnid;
}}

<? xml version="1.0" encoding="UTF-8" standalone="yes"?>

<ServiceResponse>

    <message>Database Successfully Updated</message>

    <returnid>1</returnid>

</serviceResponse>
```

Above we can see the XML representation of the Java object with JAXB annotations.

The table below contains the different end points defined in the SearchLog resource class which provides functionality for capturing the search process history as well as for retrieving this search history in order to be displayed in the user interface tools provided part of LearnWeb2.0 system.

**Table 2: Search Tracker Service Endpoints** [base URI: /searchlog]

| Endpoints | Description of functionality |
| --- | --- |
| **POST** /querylog | This method posts the user query along with the search type, session ID, user ID and timestamp. |
| **GET** /filterqueriesbytime/{start_timestamp}/{end_timestamp} | This method filters the query history corresponding to a user between two user specified timestamps. |
| **DELETE** /deleteuserqueries | It deletes a set of user queries from the table. |
| **POST** /searchcomment | It posts the user comments on a particular search process to the database. |
| **GET** /commentsbyresultsetid | It returns the set of comments that was posted |

| | for this particular search process. |
|---|---|
| **POST** /xmlresultsetlog | It stores each resource one by one to MySQL. |
| **POST** /xmlbatchresultsetlog | This method posts the resources as a batch to the database |
| **GET** /resourceurlsbyresultsetid/{resultsetid} | This method returns a list of resource URLs for a resultset_id that corresponds to a query posted at a particular timestamp. |
| **GET** /resourcesbyresultsetid/{resultsetid} | This method returns the list of resources for resultset_id for given query and timestamp. |
| **POST** /resourcelog | It logs the actions on resources such as resource click or resource saved. |
| **POST** /updateresultset | This method updates the resources table if a particular resource is saved. |
| **GET** /resourceslogbyresultsetidandaction/{resultsetid}/{action} | It returns the resource log information for a resultset_id and particular action. |
| **POST** /updateviewingtimelog | It stores the viewing time for the resource in the case of image or video search. |
| **POST** /updatebatchviewingtimelog | It logs the viewing time of resources as a batch to the database. |
| **POST** /taglist | It posts the list of tags that the user had annotated for a particular search process. |
| **GET** /tagsbyresultsetid/{resultsetid} | It returns the set of tags corresponding to a particular resultset_id. |
| **GET** /searchhistory/{userid} | It returns the raw search history for a user. |
| **GET** /searchhistorybydate/{userid} | This returns the search history grouped by date for a given user ID. |
| **GET** /searchhistorybypages/{userid}/{offset}/{limit} | This returns the search history in pages for a given user and given the offset and limit. |
| **GET** /searchhistorybyquery/{query} | This method returns a set of queries from the search history similar to the given query. |

| **POST** /shareresultset/{userx}/{usery} /{resultsetid} | This method stores the sharing of particular resultset from one user x to another user y. |
|---|---|
| **GET** /sharedresultsetsbyuserid/{userid} | This method returns the list of resultsets shared with the user given by the user ID. |

### 4.2.2 LearnWeb2.0 REST client

Jersey the reference implementation of JAX-RS (JSR 311 & 339) provides a client API to support the ease of development of REST client. It is a high level Java based API that supports the interoperability with RESTful web services and enables the development of concise and efficient reusable client system which abstracts the already existing client side HTTP implementations. Without the client API, the users would need to use low level *HttpURLConnection* to access the search tracker service endpoints where there is more focus on the client − server constraints for the exchange of messages rather than a web resource, identified by a respective URI and the use of HTTP methods to access and manipulate that resource. Thus the Jersey client API which wraps support for this low-level implementation is used to develop the REST client for LearnWeb2.0 system.

```java
URL url = new URL (http://. ../searchlog/querylog);
HttpURLConnection conn = (HttpURLConnection) url.openConnection();
conn.setRequestMethod ("POST");
conn.setRequestProperty ("Accept", "application/xml");
conn.setDoInput (true);
conn.setDoOutput (false);
BufferedReader br = new BufferedReader (new
InputStreamReader(conn.getInputStream()));
String line;
while ((line = br.readLine()) != null) {
        //. . .
}
```

The above code is implemented using *HttpURLConnection* for accessing the GET endpoint in order to retrieve the search history corresponding to a user given the ID. The same code implemented using the Jersey client API is shown below, which highlights the ease of implementation of client systems.

```
public ArrayList<HistoryByDate> getSearchHistoryByDate(int userId){

ArrayList<HistoryByDate> historyByDates = new ArrayList<HistoryByDate>();

        If (userId != -1)
        {
                WebResource web = client.resource(searchHistoryByDateURL+userId);
                ClientResponse resp =
                web.accept(MediaType.APPLICATION_XML).get(ClientResponse.class);

                if (resp.getStatus() != 200) {
                        throw new RuntimeException("Failed : HTTP error code : "
                                            + resp.getStatus());
                }

                HistoryByDateList historyByDateList =
                resp.getEntity(HistoryByDateList.class);

                historyByDates.addAll(historyByDateList.getHistoryByDates());
        }

        return historyByDates;
}
```

Initially to use the client API an instance of a client is created. The client instance could be configured by setting properties in the map returned by the getProperties method. Once the client instance is created a web resource is obtained by creating a reference to service endpoint URI such as http://localhost:8080/searchlog/searchhistorybydate/8638. As client instances are expensive resources, multiple web resources are created using the same instance as the building of responses and receiving of requests are thread safe operations. Web resource instance will utilize *HttpURLConnection* for communication with the search tracker service.

The requests to a web resource are built using a *RequestBuilder* in which the terminating method is a HTTP method as we can see in the above example it terminates with a GET request. The above request contains an accept header of application/xml or application/json, which specifies the representation of the resource that the service will return and it will accept. If the request has an entity as in the case of PUT, POST and delete then the terminating HTTP method will be declared in the call. If the response has an entity then Java type instance is declared in the HTTP method which de-serializes the response entity to that instance.

In the above code, Java type *ClientResponse* is used as the response meta-data is required which contains information of the response status, headers and the entity. From this response if the appropriate status is not returned we throw a runtime exception with the error HTTP status error code. The REST client follows similar semantics for defining methods to access the various different endpoints made available through the search tracker service.

**Table 3: HTTP Status Codes defined according to RFC2616**

| Code Name | Description |
|---|---|
| 200 OK | The request has been successfully completed. |
| 201 Created | The request has been executed leading to the creation of a new resource. |
| 202 Accepted | The request has been accepted for processing, but not yet completed. |
| 204 No-Content | The service has fulfilled the request but there is no return entity. |
| 400 Bad Request | The request has malformed syntax so it cannot be processed. |
| 401 Unauthorized | The request must have user authentication. |
| 403 Forbidden | The service understands the request but is refusing to fulfill it. |
| 404 Not Found | The requested URI is not found as an endpoint of the service. |
| 405 Method Not Allowed | The method included in the request from the client is not allowed by the resource identified by the given URI. |
| 415 Unsupported Media Type | The service is refusing to process the request as the entity is in a format not supported by the requested. |
| 500 Internal Server Error | The server encountered an unexpected error which prevented it from completing the request. |
| 503 Service Unavailable | The service is not able to handle any requests due to maintenance of the service or temporary overloading. |

### 4.2.3 Search History Modules in LearnWeb2.0

Developing with JSF2.0 has many advantages. Restriction of the processing of data to a bean eases the development of graphical user interfaces. Managed bean is a regular Java class managed by the JSF framework. The persistent values are represented as bean properties which are accessible through getter and setter methods, and the application specific logic is implemented using action controller methods defined within the bean. Managed beans works as a model for the UI components and are accessible through the JSF pages. Each managed bean has a scope annotation which indicates for how long the bean will remain in scope.

**Table 4: Managed Bean Scope Annotations**

| Scope | Description |
|-------|-------------|
| @RequestScoped | It gets created upon a HTTP request and gets destroyed when the HTTP response is sent for that associated request. |
| @ViewScoped | The bean is in scope as long as the user is interacting with the same JSF view in the browser. It gets created on HTTP request and gets destroyed when the user postback to a different view. |
| @SessionScoped | Bean lives for as long as HTTP session lives. It gets created upon the first HTTP request involving the bean and it gets destroyed when the HTTP session is invalidated. |
| @ApplicationScoped | The bean is in scope for as long as the application lives. It gets created upon the first HTTP request involving the bean or when the application starts up and is destroyed only the application is shut down. |
| @NoneScoped | Bean gets created upon an expression language (EL) evaluation and gets destroyed immediately after the EL evaluation. |

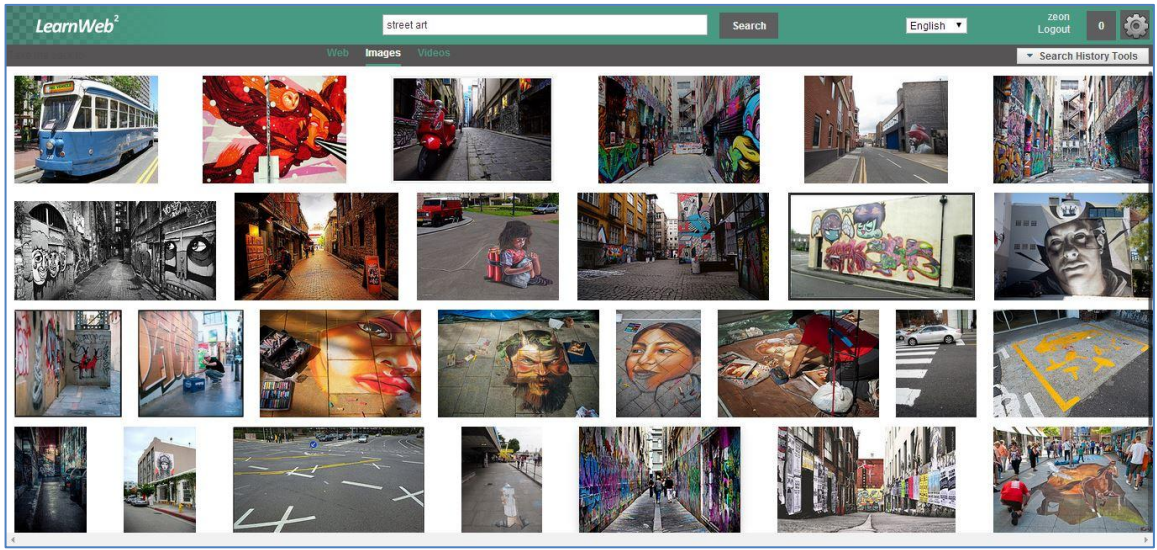

**Figure 4.6: LearnWeb2.0 Search Page**

The search page is managed by the *SearchBean* backing bean. When the user enters the keyword to be searched and clicks on the search button, the respective setter for the bean property corresponding to the search input field is called and then the action controller method to perform the search using the given keyword is executed. The search query data comprises of the query, timestamp when it was entered, type of search, session id and user id which is sent to the REST client.

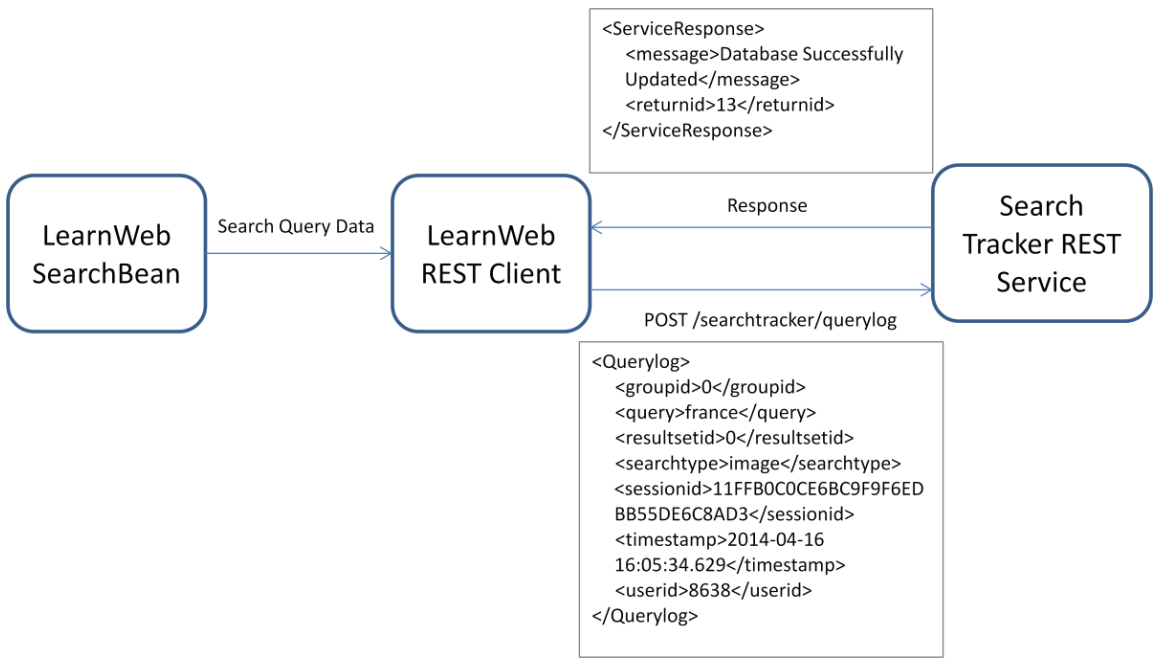

The REST client then creates a JAXB object using this query data received and sends a POST request to the respective endpoint. The JAXB object serializes the Querylog object to a XML representation which is attached to the request entity body and it is de-serialized when received by the search tracker service. This data is then stored in the corresponding table and a result set id is returned which is sent back as a response along with a message in the XML representation back to the REST client.

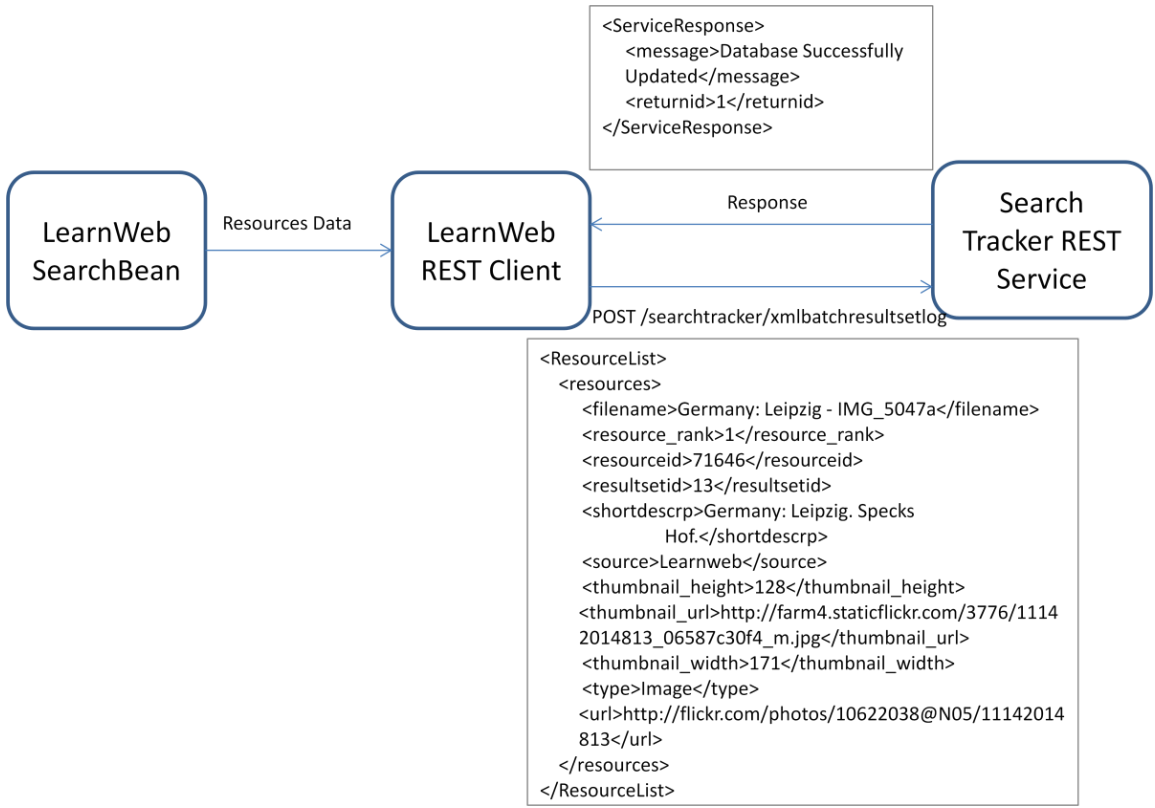

Once the query is issued, the results retrieved are then sent to the REST client where it is temporarily stored before sending it as a batch to the search tracker service using a POST request with the resources list in a XML format to the respective endpoint. The search tracker then executes a batch insert of these resources to the database and returns a response back to the REST client. The batch request is sent when either one of the cases occurs first: after a timeout of 10 minutes; when there is a query change; when the session expires. The result set id that was returned back to the client after posting the query information, which is attached to every result that is added the batch of results in order to create a relation between the query and the corresponding results.

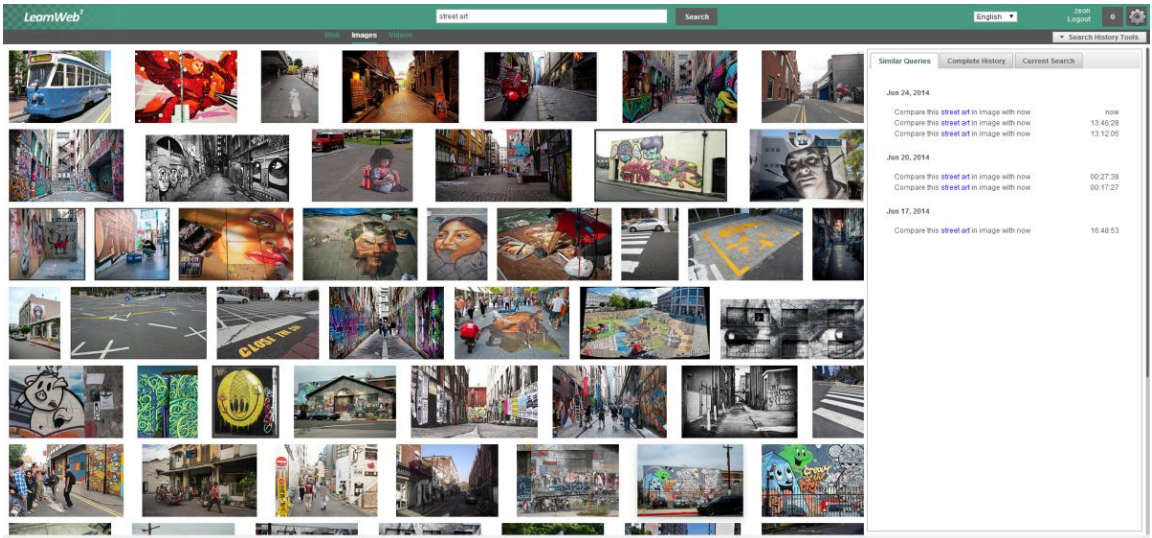

**Figure 4.7: Search Page with search history modules**

The search history modules are initially not displayed when a user visits the search page so that it doesn't confuse with the search process. When the user does want to view the search history modules, he can click on the search history tools button. The show/ hide functionality is implemented using JQuery which animates the sliding of the search history modules into the search page as well as adjusting the CSS of the element containing the results to display the view consistently. In order to keep track of the user preference for the display of search history modules, I used the JavaScript web storage feature which stores the user id and the last option chosen that is either show or hide. So the next time the user carries out a search process depending upon the last option chosen the search history module is either displayed or hidden.

The *tabView* component of the PrimeFaces library is used to implement the tab structure of the search history modules interface on the search page. Each search history module is wrapped within a *tab* component. The *tabView* component is given a widgetVar name which is used as a client side variable in JavaScript. This variable is used to keep track of which tab is currently selected in the search history view with the help of the *select* and *getActiveIndex* methods. The data for the search history modules are managed with the help of the *SearchHistoryBean*. The search history data is asynchronously loaded with the help of *remoteCommand* functionality of PrimeFaces, thus not affecting the retrieval efficiency of the search results for a particular query.

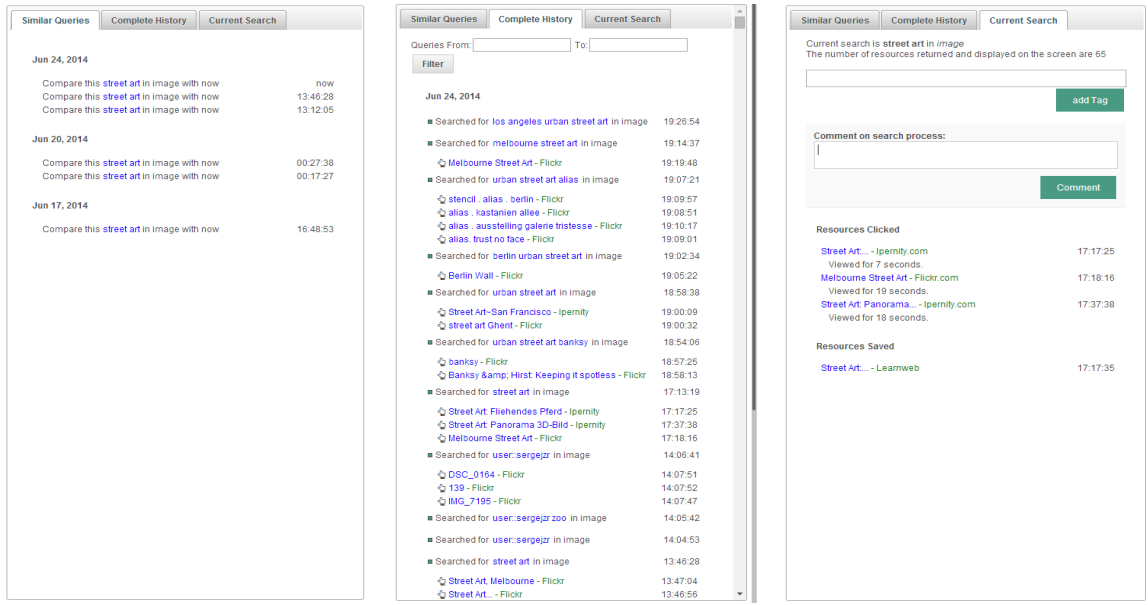

**Figure 4.8: Search History Graphical User Interfaces**

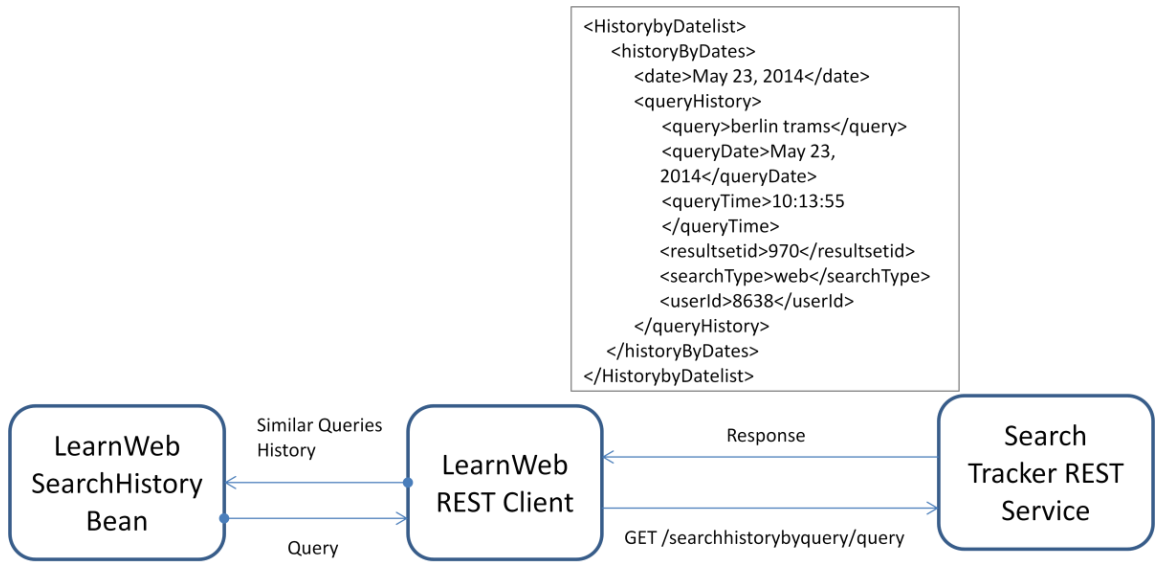

For the case of displaying the similar queries history, a request is sent to the REST client to retrieve the required data by passing the query. The REST client then issues a GET request to the corresponding endpoint with the query as a parameter. The search tracker then retrieves the query history similar to the given query from the database. This data is sent back to the client in a XML representation as a part of the response body. The client then forwards these similar queries back to LearnWeb in order to be displayed in the view. Similarly the complete history data is obtained from the search tracker.

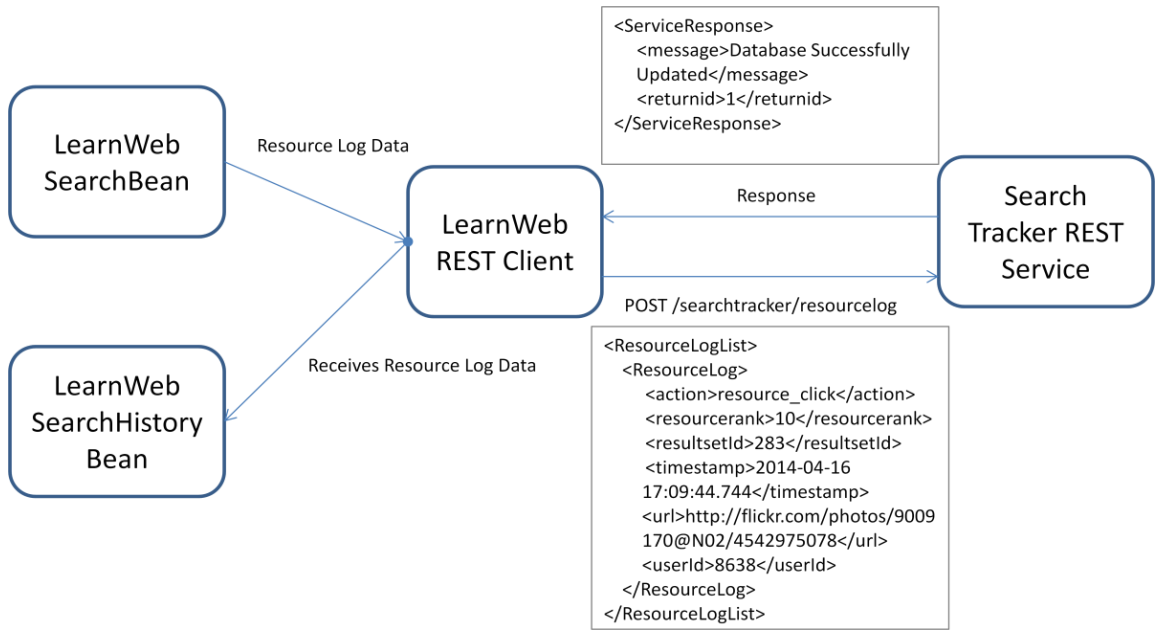

In the current search interface, the resource log actions such as resources clicked/saved, for how long the resource was viewed are displayed. The user can also document his thought process by specifying comments and tags. The resource log data is sent from the *SearchBean* and it is stored temporarily in the REST client until it is sent as a batch to the search tracker service using a POST request to the corresponding endpoint. The POST request is called after the batch of resources temporarily stored, are sent to the service. In order to display this information in the current search interface the *SearchHistoryBean* sends a request to the client to transfer the resource log data temporarily stored.

Following a similar sequence the viewing time of the resource as well as the comments and tags are displayed. The action controller methods that support the addition of comments, tags and deletion of tags are part of the *SearchHistoryBean* and not the *SearchBean*. Once a set of tags is entered for a particular search process, these set of tags are also carried forward to the next search process to model a flow similar to a search task trail as the immediate next search process could be related. The user has to manually edit the set of tags already present in the new search process, and this new updated list is then carried forward to the next. To implement this we keep the temporary storage of the tags in the client until an end in session is detected.

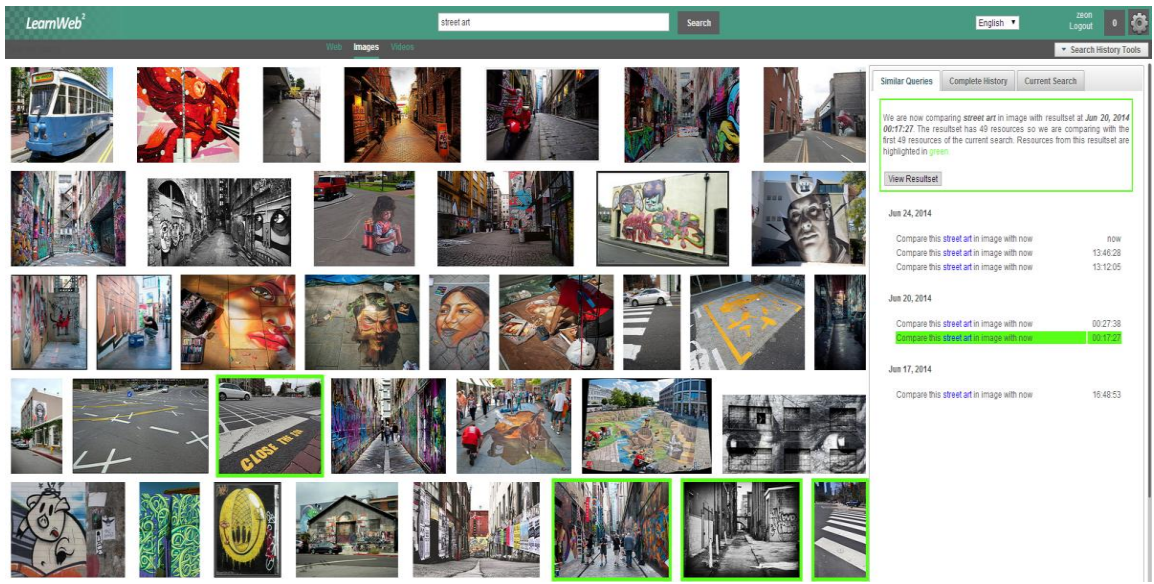

**Figure 4.9: Comparison of searches over time**

When we refer to comparison of searches over time, we mean the comparison of the first 'n' results that are currently retrieved for a particular query with the 'n' results that were displayed for a similar query that was posted earlier. The resources that were not present in the old set of resources are highlighted in green.

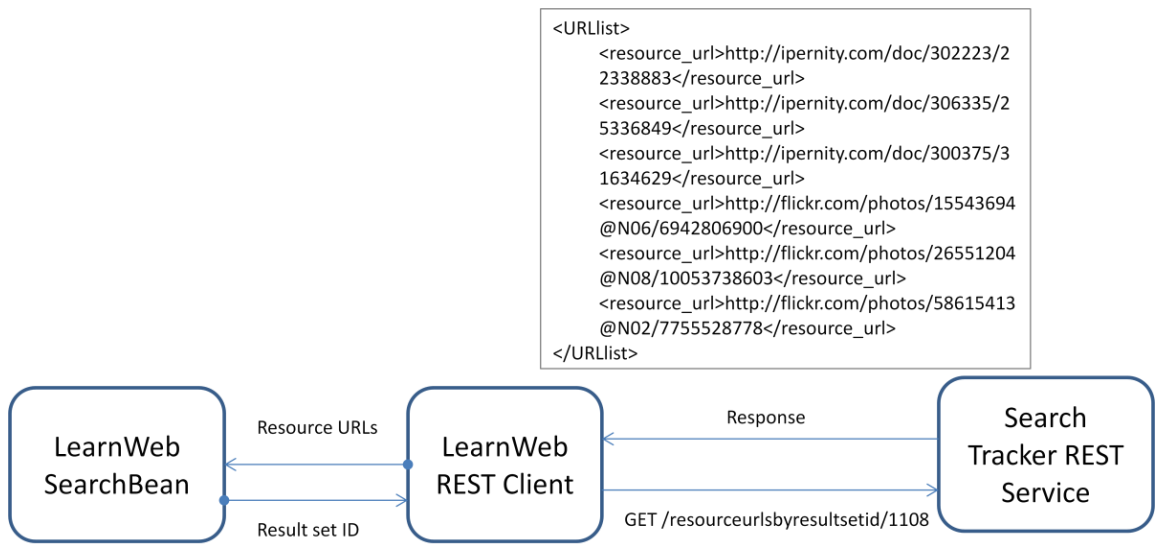

The result set ID corresponding to the selected similar query is sent from *SearchBean* to the REST client. A GET request is sent to the respective URI along with the result set ID as a parameter to obtain the resource URLs. Once the resource URLs are retrieved from the database, it is sent back to the client in the form of XML representation. The *SearchBean* then compares the set of resources currently retrieved with the set of resource URLs that were returned for the result set ID.

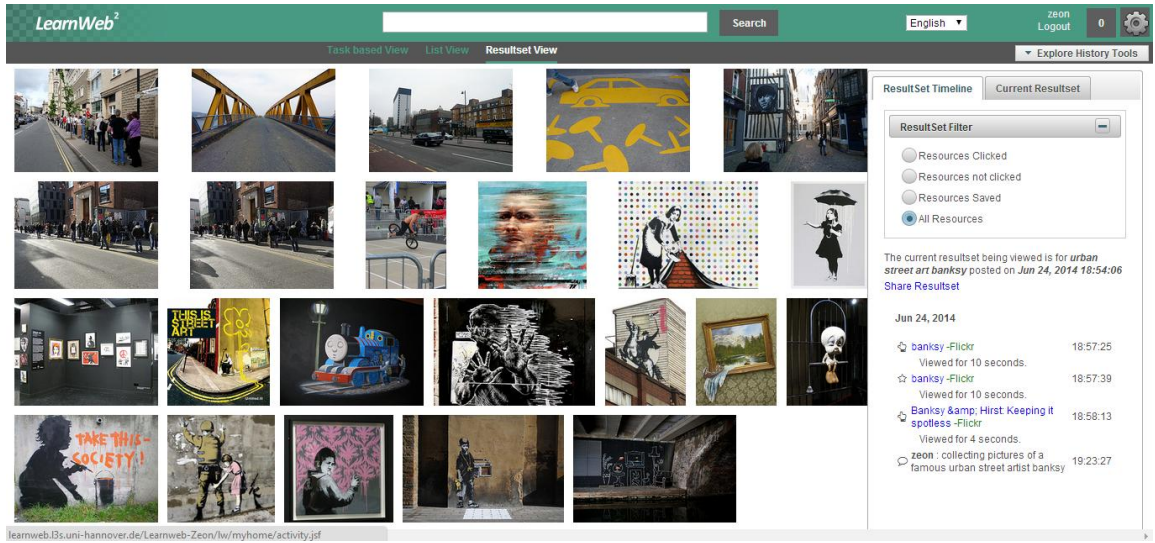

**Figure 4.10: View result set for past search context**

"View result set" for past search context provides a view of the set of resources that were returned for a particular query which was issued earlier. The set of resources is only the ones which were displayed on the screen and viewed by the user. It presents the set of resources in a grid view similar to that of the search page. The user has the flexibility to filter the set of resources viewed based on the resources clicked, not clicked and saved. There is also a timeline presenting the different user events such as resources clicked/saved, view time of a resource and comments which is shown in the right panel. The user could also save a particular resource from this set to a particular group, which helps the user capture more resources that are relevant during the second round of investigation. The current result set tab in the right panel assists the user in keeping track of the various resources he clicked or saved in the result set view. Additional comments as well as tags could also be added to this previous search context. A user could also share this set of resources along with the user actions to another user so that the other user could help collaborate and add more resources or he could carry out a review on how valid the built corpus is.

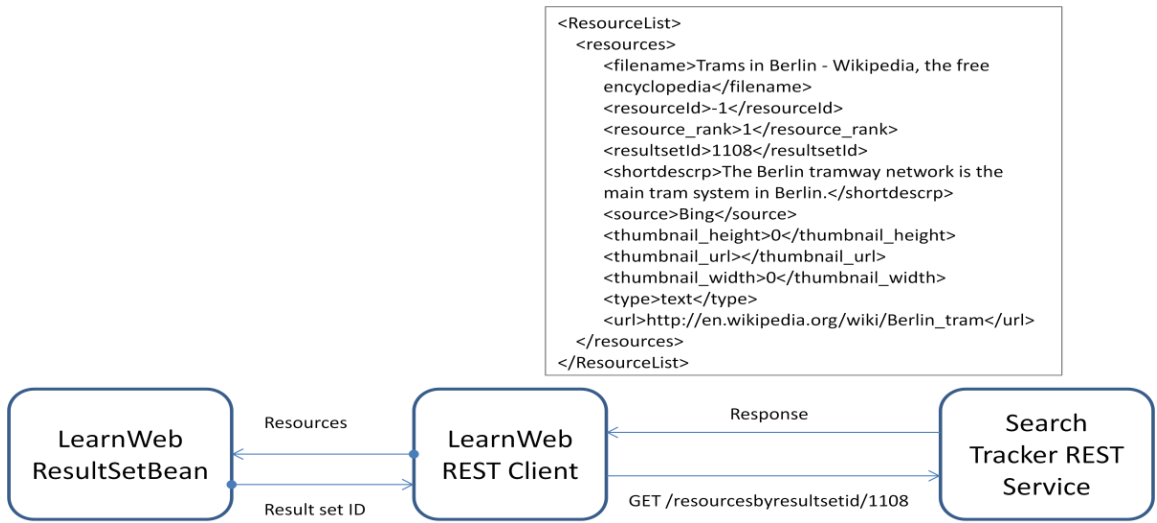

```
<ResourceList>
    <resources>
        <filename>Trams in Berlin - Wikipedia, the free
        encyclopedia</filename>
        <resourceId>-1</resourceId>
        <resource_rank>1</resource_rank>
        <resultsetId>1108</resultsetId>
        <shortdescrp>The Berlin tramway network is the
        main tram system in Berlin.</shortdescrp>
        <source>Bing</source>
        <thumbnail_height>0</thumbnail_height>
        <thumbnail_url></thumbnail_url>
        <thumbnail_width>0</thumbnail_width>
        <type>text</type>
        <url>http://en.wikipedia.org/wiki/Berlin_tram</url>
    </resources>
</ResourceList>
```

The result set ID corresponding to the selected query is sent from *ResultSetBean* to the REST client. A GET request is sent to the respective URI along with the result set ID as a parameter to obtain the set of resources that were displayed on the screen for that query. Once the resources are retrieved from the database, it is sent back to the client in the form of XML representation. The *ResultSetBean* then wraps the set of resources retrieved in a format similar to that of the search page, so that the view will be identical to it. In order to filter the set of resources by user actions, that is, resources clicked/saved and resources not clicked, the client sends the request to the same URI using an additional parameter which is the user action.

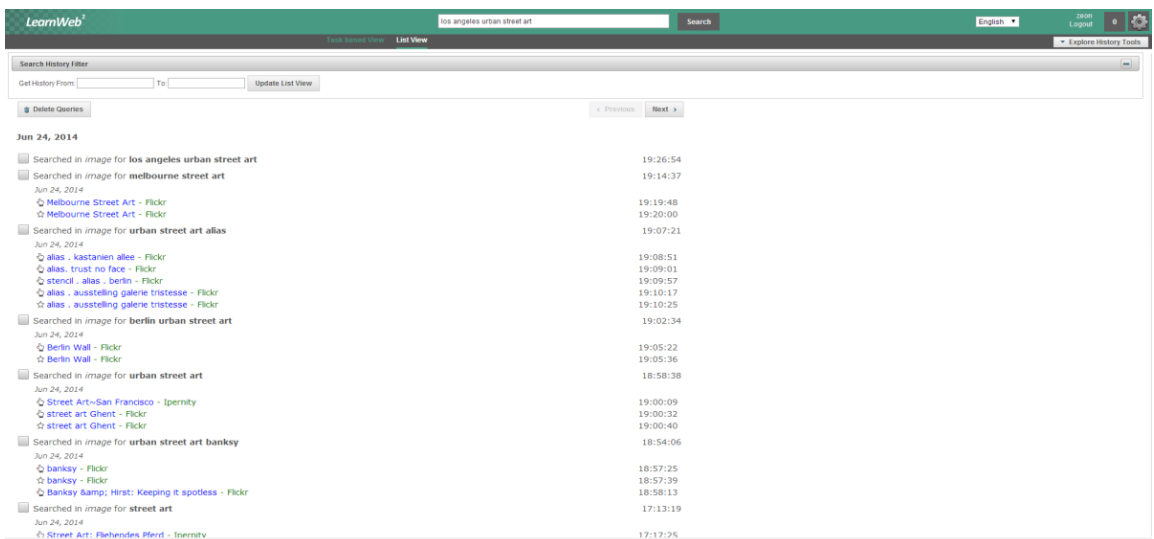

**Figure 4.11: Explore Search History Page**

In the explore search history page, the user can see the queries issued along with the results clicked or saved by the user displayed in reverse chronological order. If certain search contexts are not relevant to the user, he could delete those contexts. The search history is displayed as pages so that it will be displayed quickly and the user can retrieve the other pages as and when it is required. The user could also filter the search history by time. On hovering over the queries issued in the search history, a view result set button appears, which on clicking redirects the user to the page which displays the set of resources that were returned for that query.

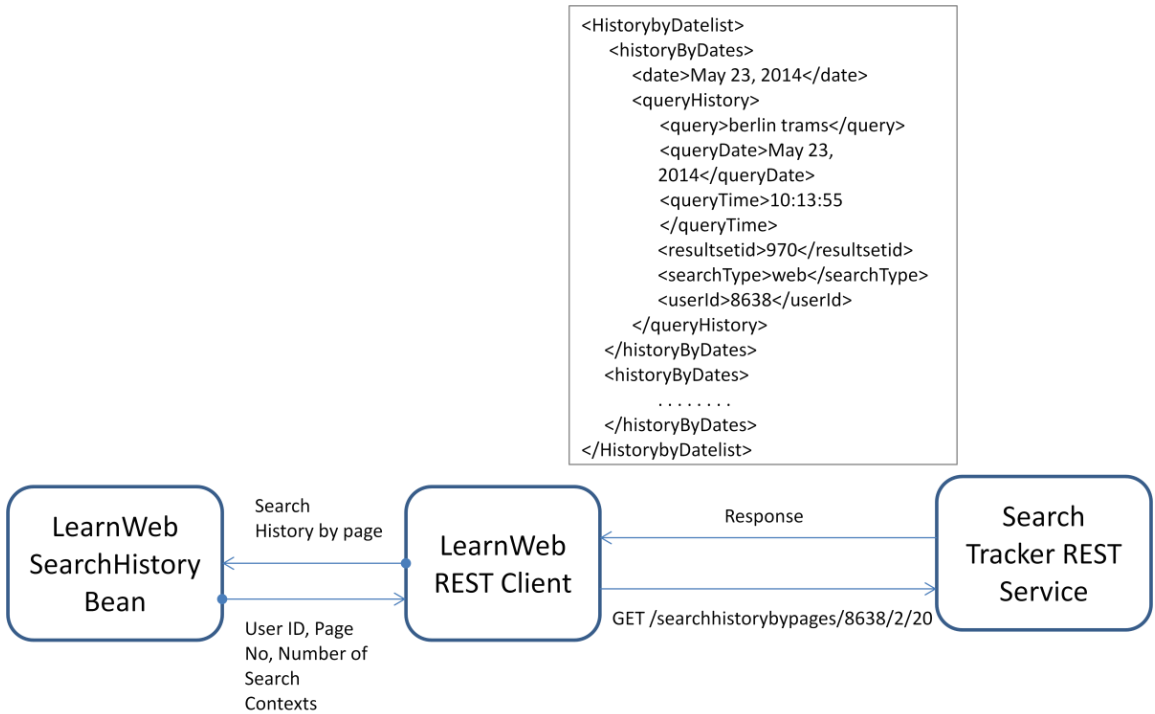

The user ID, page number of search history and number of search contexts is passed from the *SearchHistoryBean* to the REST client. A GET request is sent to the respective URI along with the user ID, page number and number of search contexts as parameters to obtain the set of search contexts. Once the search contexts are retrieved from the database, it is sent back to the client in the form of XML representation. The *SearchHistoryBean* then displays the set of search contexts in a list view similar to the layout shown in figure 4.11.

## 4.3 Unit Test Cases

Unit testing is a software testing method through which different modules of the system are tested to determine if they are fit for use. Unit test cases highlight the characteristics that are critical for the success of a module/unit.

**Table 5: Unit Test Cases of implemented system**

| Test Case ID | Description | Expected Outcome | Actual Outcome | Result |
|---|---|---|---|---|
| 1 | Capturing the search context data and sending it to the REST client | Search Context data stored in lists | Search Context data stored in lists | Success |
| 2 | REST client communicating search context data to search tracker REST API | Search Context data successfully sent | Search Context data received by REST API | Success |
| 3 | Search tracker REST API storing search context data in MySQL Database | Search Context data stored in database | Search Context data stored in database | Success |
| 4 | Retrieving search context data from search tracker REST API | Search Context data retrieved from database | Search Context data retrieved from database | Success |
| 5 | Display of search context data in LearnWeb2.0 platform | Search context data displayed | Search context data displayed | Success |

# CHAPTER 5

# RESULTS AND DISCUSSION

The system was successfully implemented and integrated into the LearnWeb2.0 platform. All the functionalities that were specified have been incorporated into the system. LearnWeb2.0 users can now use the search history modules to help understand how a multimedia corpus was built. These search history modules help visualize the search contexts that have been captured before.

Only those users who belong to a course that has search context capturing and visualization enabled can use the modules that have been integrated for building multimedia corpora. While carrying out a search for the first time using LearnWeb2.0 the search history modules are not displayed, but if the user needs to see these modules it can be accessed using the search history tools button in the search page. And this preference of search history modules being displayed is saved for the user, so that the next time the user searches for something depending upon his last preference the search history modules are either displayed or not.

In the search page, as the user carries out a search the similar queries tab part of the search history modules displays the same queries that were issued by the user earlier. The user can carry out a comparison of the set of results that are returned now with the set of results that were returned earlier for a similar query issued. Then the results from the current set which was not present in the earlier set of results are highlighted separately giving the user an idea of how the corpus corresponding to a query has changed. The user can also keep track of his actions on the search page using the current search tab, which displays the resources clicked/ saved and for how long they were viewed. Documentation of a particular search could also be done using comments or tags.

A detailed search history page displays the queries issued by the user, along with the corresponding resources that were clicked or saved in reverse chronological order. Using this view the user can gain perspective of the steps that were taken in order to build a multimedia corpus. Most of the modern services these days don't stream all the results at once, so that the system is responsive as the user may not wait till all the results are loaded. Thus the search history is displayed as pages allowing the user to only view a small set of results initially and displaying the next set of pages depending upon their request. The user can delete certain search contexts from the search history. The search history can also be filtered between a set of dates specified by the user.

In order to understand from which set of resources a particular result was chosen for the multimedia corpus, the user can also revisit the set of resources corresponding to the query from which the result was chosen. The set of resources are displayed in a view similar to that of the search page, and the user can filter this set depending on user actions such as resources clicked, not clicked or saved. This helps the user carry out a detailed analysis of the set of resources and add more results to the multimedia corpus if needed or provide additional comments or tags.

Multiple people can build a multimedia corpus together, thus the system allows this kind of collaboration by sharing of a set of resources corresponding to a query posted earlier. Other users could also review the multimedia corpus which was built by looking at the set of resources and the results that was chosen by a user to understand if the choices made were biased or not.

This implementation was carried out using different technologies such as JavaServer Faces 2.0 (JSF2.0), Jersey API, JAX-RS API, PrimeFaces, MySQL and Apache Tomcat. The performance and efficiency of the system was noted as well as analyzed, and it showed a satisfactory outcome.

# CHAPTER 6

# CONCLUSIONS AND FUTURE ENHANCEMENTS

The thesis describes the importance of capturing search contexts and how it could be supportive for building multimedia corpora. It also describes the importance of capturing search contexts which will be helpful in resuming an interrupted search task or for the management of complex search tasks which spans over multiple days.

Next, it talks about the related work for capturing search contexts how they are split into three approaches: link, page and search centric. Of these approaches the search centric is best suited for the scenario of providing support for building multimedia corpora. A set of existing systems implemented to capture search contexts are discussed along with their drawbacks and how they are different from the search tracker REST service implemented part of my thesis. The LearnWeb2.0 multimedia search system along with REST web services is also described.

Thirdly, the thesis describes the analysis and design of the project. The non-functional and functional requirements the system must have are specified. The system design is covered in detail comprising of the REST architectural style and the LearnWeb2.0 system architecture. Moreover, it is accompanied by behavioral diagrams in the form of use case diagrams and logical flow of the system through the use of sequence diagrams.

Furthermore, the thesis explains the tools and the implementation strategies that have been used to build the required system. This comprises of a detailed presentation of the different implementation steps that is the search tracker REST service used to capture search contexts, the REST client for LearnWeb2.0 in order to communicate with the search tracker REST service and finally the search history graphical user interfaces integrated into LearnWeb2.0 to visualize the saved search contexts.

All steps are described clearly along with descriptive diagrams, flow diagrams, code snippets and screen shots of the actual system which was implemented.

While the system has been implemented, there are many further enhancements we could add to the system. First of all we could implement similarity measures such as Levenshtein distance, Jaccard similarity or Hamming distance instead of using exact string match for finding the similarity between queries which are to be displayed in the similar queries tab part of the search history graphical user interface.

In addition, we could index the comments and tags given by the user for search contexts and then provide search functionality within the explore search history page. This will help the user in easily accessing only the search contexts he wants to observe and thereby helping him understand the various steps that were already taken to build a particular multimedia corpus.

Finally, implementing an automatic search task detection algorithm over all the search contexts captured to categorize them into different tasks. One of the approaches to do this has been discussed in [22] where the query similarities was measured using time and word based features to classify if two queries belong to the same task, and then a clustering approach is implemented to merge similar queries into the same task using a Support Vector Machine (SVM) classifier. Once these tasks have been identified, we could represent this in a collapsible tree layout in d3js where the user clicks on node representing a search task and it shows the set of queries corresponding to that search task. This view will be more intuitive for the user to understand how a multimedia corpus was built.